\documentclass[limeno]{jfm}

\usepackage{physics}
\usepackage{graphicx}

\usepackage{newtxtext}
\usepackage{newtxmath}
\usepackage{natbib}
\usepackage{hyperref}
\hypersetup{
    colorlinks = true,
    urlcolor   = blue,
    citecolor  = black,
}

\newcommand{\RomanNumeralCaps}[1]
\linenumbers

\newcommand\Oh{\textrm{Oh}}
\newcommand\Bo{\textrm{Bo}}
\newcommand\We{\textrm{We}}

\newcommand\force{\mathcal{T}}
\newcommand\R{R_0}

\newcommand{\avg}[1]{\langle #1 \rangle}

\title{Bubble shape oscillations in a turbulent environment}
\author{Ali\'enor Rivi\`ere\aff{1},
\corresp{\email{alienor.riviere@espci.fr}}
Kamel Abahri\aff{1}
\and St\'ephane Perrard\aff{1}}
\affiliation{\aff{1}PMMH, CNRS, ESPCI Paris, Universit\'e PSL, Sorbonne Universit\'e, Universit\'e de Paris, 75005, Paris, France}

\begin{document}
\maketitle
\begin{abstract}
    We investigate bubble deformations in an homogeneous and isotropic turbulent flow by means of direct numerical simulations of a single bubble in turbulence.
    We examine interface deformations by decomposing the local radius into the spherical harmonics base.
    We show that the linear dynamics of each mode, (for low Weber number), can be modeled by a forced stochastic linear oscillator.
    We measure the coefficients of the model directly from the modes' statistics.
    We find that the natural frequency corresponds to the Rayleigh frequency, derived in a quiescent flow.
    However, dissipation increases by a factor 15 compared to the quiescent case, at $\Re_\lambda = 55$.
    This enhanced dissipation originates from a thick boundary layer surrounding the bubble.
    We demonstrate that the effective forcing, originating from the integration of pressure over the bubble surface, is independent on bubble deformability.
    Therefore, the interface deformations are only one-way coupled to the flow.
    Eventually, we investigate the pressure modes' statistics in the absence of bubbles and compare them to the effective forcing statistics.
    We show that both fields share the same pdf, characterized by exponential tails, and a characteristic timescale corresponding to the eddy turnover time at the mode scale.
\end{abstract}

\section{Introduction}
\subsection{Broad context}

Bubbly turbulent flows are widely used in industrial processes to enhance mass transfers and chemical reactions.
Examples involve bubble column reactors \citep{risso2018} and emulsifiers \citep{haakansson2019,haakansson2021} for instance.
In geophysical contexts, bubbles are known to control aerosol productions at the ocean-atmosphere interface, while playing a major role in gas transfers \citep{deike2022}.
In both industrial and natural situations, the knowledge of the bubble size distribution and its temporal evolution is necessary to predict mass transfers across bubble interfaces.
As a consequence, the study of bubble breakup in turbulence has received considerable attention since the pioneer works of \citet{kolmogorov1949} and \citet{hinze1955}.
They predicted that, for bubbles of size lying within the inertial range of the turbulent cascade, bubble dynamics and breakup are primarily controlled by the balance between inertial and capillary forces. 
This ratio defines the Weber number $\We(d) = \rho U^2 d/\gamma$, where $\rho$ is the liquid density, $U$ a characteristic velocity, $d$, the bubble volume equivalent diameter and $\gamma$ the surface tension between gas and liquid.
In turbulence, assuming bubbles break due to velocity fluctuations at their scale, Kolmogorov and Hinze postulated that the characteristic velocity $U$ is the average velocity increment at the bubble scale $\langle \delta u^2(d) \rangle^{1/2}$.
When kinetic and capillary forces balance we have $\We(d_h) \approx 1 \approx \We_c $, which defines a critical size, the Kolmogorov-Hinze scale $d_h$ separating statistically stable bubbles ($d<d_h$) from unstable bubbles ($d>d_h$).

However, the main physical mechanism leading to breakup remained to be understood.
\citet{sevik1973} proposed a resonant mechanism, in which bubble breaks due to series of excitation at its natural frequency, while other authors argue that large fluctuations are necessary for a bubble to break \citep{lee1987,luo1996,wang2003}. 
To address this question, several authors  describe bubble deformation dynamics, either with the help of a linear damped harmonic oscillator \citep{risso1998,ravelet2011,masuk2021simultaneous} on the bubble Rayleigh modes \citep{rayleigh1879}, or via a tensorial equation for the main bubble axis of deformations \citep{masuk2021model}.
The latter assumes that bubble shape is mostly ellipsoidal while the former allows any bubble shape and describes each mode dynamics.

More generally, bubble deformations in turbulence are one over many examples of the interaction between a turbulent flow and a deformable object.
From plants oscillations in the wind \citep{langre2008}, to disks \citep{verhille2022} and fibers deformations in water \citep{rosti2018,brouzet2021}, many studies have aimed at finding a reduced dynamics for the amplitude of the relevant spatial modes of deformations, in the form of a damped harmonic oscillator, randomly forced by turbulence. A usual approach is to model the coefficients of an ordinary differential equation, as well as the statistics of a random forcing term that model the turbulent forcing.
Here, we propose to directly measure these coefficients and the forcing statistics from the deformation statistics of bubbles in turbulence, by performing numerical experiments.

We first review the bubble oscillation dynamics in quiescent flows and their phenomenological extensions to turbulent flows.

\subsection{Bubble dynamics in quiescent flows}

\citet{rayleigh1879} investigated the oscillation dynamics of inviscid drops in vacuum and bubbles in a quiescent inviscid flow. In the linear limit of deformation, the local radius of a bubble (or a drop) can be decomposed into axi-symmetric modes using the basis of Legendre functions, which are indexed by the integer $\ell \in [2, \infty]$.
Rayleigh showed that the amplitude of each mode $\ell$ follows an harmonic oscillator equation, with a characteristic natural frequency writing, 
\begin{equation}
    \omega_\ell^2 = 8(\ell -1)(\ell +1 )(\ell + 2) \frac{\gamma}{\rho d^3}
    \label{eq:RayleighNatFreq}
\end{equation}
for bubbles, with $f_\ell = \omega_\ell/(2\pi)$ the characteristic frequency.
Latter on, \citet{lamb1932} extended this work to gas bubbles oscillating in a liquid of low kinematic viscosity, $\nu$.
He found that bubbles' modes oscillate at the Rayleigh frequency with a damping rate $\lambda_\ell$, which reads
\begin{equation}
    \lambda_\ell = 8(\ell + 2)(2 \ell + 1)\frac{\nu}{d^2},
    \label{eq:LambDamping}
\end{equation}
for bubbles of negligible inertia and viscosity. In three dimensions, bubble shape can be decomposed into the real spherical harmonics base, $Y_\ell^m(\theta, \phi)$, indexed by $\ell \in [2, \infty]$ and $m$ an integer ranging from $-\ell$ to $\ell$, where $\theta$ and $\phi$ are the co-latitude and longitudinal angles in spherical coordinates.
The axi-symmetric modes of Rayleigh and Lamb correspond to $m=0$.
We denote the dimensionless amplitude of the modes in the spherical harmonics base by $x_{\ell, m}$.
The dynamics found by \citet{lamb1932} and \citet{rayleigh1879} applies to each spherical harmonics mode amplitude $x_{\ell, m}$, so that they follow a damped harmonic oscillator equation with natural frequency $\omega_\ell$ and damping rate $\lambda_\ell$ independent of $m$,
\begin{equation}
    \ddot x_{\ell, m} + \lambda_\ell \dot{x}_{\ell, m} + \omega_\ell^2 x = 0.
\end{equation}
When time is made dimensionless using the natural frequency $\omega_\ell$, this equation reads, 
\begin{equation}
    x_{\ell, m}^{\prime \prime} + p(\ell)\Oh \,{x}_{\ell, m}^\prime + x = 0
    \label{eq:oscillatorquiescent}
\end{equation}
where $p(\ell) = 2\sqrt{2}(\ell + 2)(2\ell+1)/[(\ell-1)(\ell +1)(\ell +2)]^{1/2}$ and, $^\prime$ stands for derivative with respect to the dimensionless time $\omega_\ell t$. The Ohnesorge number $\Oh = \mu/\sqrt{\rho \gamma d}$ with $\mu = \nu \rho$ compares viscous to capillary effects, and controls the quality factor $Q_\ell = \omega_\ell/\lambda_\ell \sim \Oh \, \ell^{-1/2}$ of the Lamb oscillations.

To estimate the damping rate of small oscillations, \citet{lamb1932} computed the velocity gradients of the irrotational inviscid velocity field. 
Doing so, he underestimated the dissipation rate, as shown latter by \citet{miller1968}, as most of the dissipation takes place within the bubble boundary layer, even when viscosity is low.
Another approach is given by the normal-mode analysis \citep{chandrasekhar1959,reid1960,chandrasekhar2013}, for the spherical harmonics modes. 
This theory predicts an evolution of the bubble natural frequency and damping rate with the Ohnesorge number. No explicit formulation can be however derived: one needs to solve a characteristic equation for each value of Oh.
This approach correctly takes into account viscous effects but only holds at long times, presumably when oscillations have already been completely damped, and do not describe transient dynamics.
\citet{miller1968} demonstrated that, in the limit of vanishing viscosity, the normal-mode solution converges to the irrotational one in the bubble case.
For drops, the same demonstration has been done by \citet{chandrasekhar1959} and \citet{reid1960}.

Latter on, \citet{prosperetti1977,prosperetti1980} unified the two approaches by studying the initial-value problem of a drop or a bubble oscillating is an initially quiescent flow.
He demonstrated that, regardless of the value of Oh, the damped harmonic oscillator dynamics of \citet{lamb1932} holds at short times compared to the viscous timescale, $t \ll \R^2/\nu$, where $\R = d/2$ is the bubble equivalent radius.
On the other hand, the normal mode description of \citet{chandrasekhar1959} holds at long times, $t \gg \R^2 /\nu$.
At intermediate timescales, \citet{prosperetti1977,prosperetti1980} demonstrated that the dynamics is more complex due to the existence of a memory term in the equation of motion of the modes, which couples the dynamics with the past evolution.

\subsection{Bubbles deformations in turbulence}

For a bubble immersed in a turbulent flow, additional dimensionless parameters may control the deformation. 
Let us consider a bubble of negligible inertia and viscosity, equivalent diameter $d$, immersed in a fluid of density $\rho$, dynamic viscosity $\mu$, with surface tension $\gamma$.
When the surrounding flow field is an homogeneous and isotropic turbulent flow, characterized by an energy dissipation rate $\epsilon$, and an integral length scale $L_{int}$, the Buckingham's $\Pi$ theorem predicts that the dynamics is controlled by three dimensionless numbers.
Choosing a set of dimensionless numbers which decouples viscous effects from capillary effects, we obtain that a generic measure of shape deformation $\delta$ can be written as, 
\begin{equation}
    \frac{\delta}{d} = F \left (\We(d), \Re(d), \frac{d}{L_{int}} \right).
\end{equation}
where $F$ is a dimensionless function. The Weber number $\We(d) = 2 \rho \epsilon^{2/3}d^{5/3}/\gamma$ compares inertial and capillary forces at the bubble scale. The Reynolds number $\Re(d) = \sqrt{2}\rho\epsilon^{1/3}d^{4/3}/\mu$
balances inertial and viscous forces at the bubble scale.
Finally, the ratio $d/L_{int}$ is the scale separation between the bubble scale and the integral length scale.
Note that using $\epsilon$ and $d$ we can define a characteristic velocity $U = \sqrt{2} (\epsilon d)^{1/3} = \langle \delta u^2 (d)\rangle^{1/2}$, the velocity increment at the bubble scale in homogeneous and isotropic turbulence \citep{pope2000}.
When the bubble size lies within the inertial range of the turbulent cascade, the surrounding flow is scale invariant and we expect the dynamics to be independent of $d/L_{int}$. The bubble dynamics will then be primarily controlled by the Weber number.
In the presence of gravity $g$, one must also include the Bond number $\Bo = \rho g d^2/\gamma$, comparing gravity to capillary effects. For simplicity, we will not consider gravity in this study.
This assumption is valid for bubble diameter smaller than the capillary length $\sqrt{\gamma/(\rho g)}\sim 2 \textrm{mm}$. In practice, looking at temporal evolution of bubble deformation, our model may describe shape oscillations slightly above the capillary length.

In this work we focus on bubbles which do not break, corresponding to a bubble size $d$ within the inertial range of the turbulent cascade and $d<d_h$.
For a typical turbulent flow with $\epsilon = 1 \textrm{ m}^2 \textrm{s}^{-3}$, and $\We_c \approx 3$, $d_h = (\We_c \gamma/(2 \rho \epsilon^{2/3}))^{3/5} \approx 8 \textrm{ mm}$ and $\Re(d_h) \approx 2300$.
Note that $\Re(d_h) \sim \rho^{1/5}\gamma^{4/3}/(\epsilon^{1/5}\mu)$ decreases as $\epsilon$ increases for a given pair of liquid-gas. It is worth mentioning that, as a consequence, an increase of the global Reynolds number of the flow induces more viscous effects at the Hinze scale. 

In order to predict bubble breakup, \citet{risso1998} introduced a forced linear damped oscillator equation to describe the dynamics of sub-Hinze bubbles. Observing that the average deformation increases linearly with We, up to the threshold for bubble breakup, they postulated that a linear dynamics would be valid to describe bubble deformations up to this point.
They assumed that the deformed radius $R(t)$ evolves following
\begin{equation}
    \ddot R + \lambda \dot R + \omega^2R = F_{ex}(t)
    \label{eq:modelRisso1}
\end{equation}
where $\lambda$ is a damping rate, $\omega$ a natural frequency and $F_{ex}(t)$ an instantaneous forcing from turbulence.
Bubble deformations and breakup are mainly controlled by the second spherical harmonics modes $\ell=2$, which correspond to oblate-prolate oscillation \citep{risso1998,ravelet2011,perrard2021}.
As a consequence, as a first guess, they used the Rayleigh natural frequency of mode 2, $\omega = \omega_2$, equation~\eqref{eq:RayleighNatFreq}, and the Lamb damping rate $\lambda = \lambda_2$, equation~\eqref{eq:LambDamping}, even though these values only hold in a quiescent irrotational flow.
Then, following the original idea from \citet{kolmogorov1949} and \citet{hinze1955}, they assumed that the turbulent forcing from turbulence scales as the square of the instantaneous velocity increment at the bubble scale $\delta u(d, t)^2$, leading to a forcing $F_{ex}(t) = Kd\delta u(d, t)^2 $ from dimensional analysis, where $K$ is a numerical constant of order 1.
Doing so, they assumed that the presence of the bubble does not strongly affect the flow properties, so that the flow statistics correspond to the single fluid case.
Expressing length in units of $d$, and time in units of $1/\omega_2$, equation~\eqref{eq:modelRisso1} is now written as
\begin{equation}
    r^{\prime \prime}+ 20 \sqrt{2/3}\,\Oh \,r^{ \prime}+ r = \tilde{K} \We(t)
\end{equation}
where $\tilde{K}$ is also a constant of order 1 and $\We(t) = 2 \rho \delta u(d, t)^2 d/\gamma$ is the instantaneous bubble Weber number.
This model is essentially the same as equation~\eqref{eq:oscillatorquiescent}, with an additional random forcing term.
This equation has been widely used for bubbles \citep{ravelet2011,lalanne2019,masuk2021model,masuk2021simultaneous} and drops \citep{galinat2007,maniero2012,haakansson2021,roa2023} oscillations in turbulence with the adequate expressions of the damping rate and natural frequency.

However, there is no guarantee that the bubble natural frequency and damping rate remain unchanged compared to the quiescent case.
They may \textit{a priori} depend on both $\Re$ and $\We$.
Indeed, surrounding flows are known to modify the natural frequency and the damping rate.
For instance, for bubbles in a uniaxial inviscid straining flow, \citet{kang1988} showed that the flow couples modes $\ell = 2$ and 4, inducing a reduction linear in We, of the mode $\ell =2$ natural frequency at linear order.
\citet{riviere2023} investigated numerically the deformation dynamics of bubbles in a uniaxial straining flow at large but finite Reynolds number.
Together with the linear We-dependency, they reported an additional Re-dependency of the natural frequency of mode $\ell=2$.
In addition, in inertial flows, bubble deformations are primarily driven by Eulerian pressure increments at the bubble scale \citep{qureshi2007}, which do not share the same statistics than velocity increments squared.

\subsection{Outline of the present work: Infer bubble deformations dynamics from data}

In this paper, following \citet{risso1998}, we assume a linear damped oscillator equation with a stochastic forcing for the oscillations of each mode of bubble deformation.
However, we do not presume any values for the coefficients of equation~\eqref{eq:modelRisso1} and the form of the forcing.
Instead, we directly measure from the deformations dynamics, the effective natural frequency and damping rate and compare them to the quiescent values.
We then deduce the statistical properties of the effective forcing.
To identify the origin of the effective forcing, we study the statistics of the pressure field evaluated on a sphere of bubble radius $\R$.
Eventually, we investigate the flow structure around bubbles and the local dissipation rate to discuss the origin of bubble dynamics dissipation in turbulent flows.

\section{Bubble deformations in HIT}
\subsection{Numerical set-up: DNS of a single bubble in HIT}

We perform direct numerical simulations of an incompressible gas bubble immersed in an homogeneous and isotropic turbulent flow of an incompressible liquid, using the open-source software Basilisk (\href{http://basilisk.fr}{http://basilisk.fr}) \citep{popinet2003,popinet2009,basiliksurfacetension}.
Density and viscosity ratios are set to $850$ and $25$, respectively, close to air-water ratios.
The simulation goes in two steps.
We first create an homogeneous isotropic turbulent flow by solving the one phase incompressible Navier-Stokes equations with an additional forcing term proportional to the velocity \citep{rosales2005}.
After a transient regime, the flow reaches a statistically stationary homogeneous and isotropic turbulent state.
The turbulent fluctuations are characterized by the Taylor Reynolds number $\Re_\lambda$, defined at the correlation length of velocity gradients, namely the Taylor micro-scale $\lambda = \sqrt{15\nu/ \epsilon}\, u_{rms}$~\cite{pope2000}, where $u_{rms}$ is the root mean square of the velocity.
The Taylor Reynolds number of the flow is $\Re_\lambda = u_{rms}\lambda/\nu = 55$.
We then extract snapshots of the flow and use them as flow initial conditions for bubble injection. Snapshots are separated by at least $6\,t_c$ to make sure initial conditions are independent.
The spherical bubble is injected at the center of the simulation box by changing locally the density and viscosity.
The bubble size is chosen so that it lies within the inertial range of the turbulent cascade where the flow is scale invariant. The bubble diameter to box length ratio is $0.13$.
During this second stage, forcing is maintained to sustain turbulence, but only in the liquid phase to guaranty that bubble deformations only come from the fluid forcing.
In both steps, we use adaptive meshgrid refinement in order to save computational time while resolving all the physical length scales of the problem. The minimum grid size corresponds to 34 points per bubble radius.
Details of the numerical set-up as well as a convergence study can be found in \cite{riviere2021}.

In this study we keep the flow Reynolds number constant and we vary the bubble Weber number by changing the value of the surface tension coefficient.
The bubble Reynolds number is $\Re(d) = 124$.
We explore eight values of $\We$ ranging from $\We_c \approx 3$ to $0.1\We_c$.
For each Weber number, we run between 3 and 5 simulations.
Except when the bubble breaks (at $\We=2.9$), we run every simulation for at least $20\, t_c$, so that the total time per ensemble is about $100\, t_c$.
Table~\ref{tab:NbrSimu} summarizes the exact number of simulations and total computational time per Weber number we perform.

\begin{table}
    \centering
    \begin{tabular}{|c | c c c c c c c c |} 
    \hline
     We & 2.9 & 2 & 1.43 & 1 & 0.71 & 0.46 & 0.36 & 0.27 \\ 
     \hline
     N & 5 & 5 & 3 & 3 & 5 & 3 & 3 & 5\\ 
     \hline
     $T_{tot}/t_c(d)$ & 62 & 126 & 94 & 94 & 156 & 94 & 84 & 94 \\
     \hline
     \end{tabular}   
     \caption{Number of simulations and total simulated time per values of the Weber number.}
    \label{tab:NbrSimu}
\end{table}

\subsection{Modes of deformations}

To quantify bubble deformations, we decompose the local bubble radius $R$ into the real spherical harmonics base $Y_{\ell}^m(\theta, \phi)$, where $\ell$ and $m$ are the principal and secondary numbers respectively, and $\theta$ and $\phi$ the co-latitude and longitude,
\begin{equation}
    R(\theta, \phi, t) = \R\big[1 + \sum_{\ell=2}^{\infty} \sum_{m=-\ell}^\ell x_{\ell, m}(t)Y_\ell^m(\theta, \phi)\big],
\end{equation}
and we track the modes' amplitude $x_{\ell, m}$ over time. Bubble shape is described in the bubble frame of reference so that all harmonics $\ell = 1$, corresponding to bubble translation, are null by definition.
Numerically, the bubble center is determined at each time step recursively by moving the frame origin to minimize the amplitudes of all modes $\ell=1$.
The recursion stops when the center displacement between two steps is less than $2.5.10^{-6}\R$.
This condition ensures that $|x_{1, m}|<4.10^{-6}$ for all values of m.
Note that the spherical harmonics decomposition holds as long as the local radius $R$ is mono-valued.
The procedure to compute the spherical harmonics is described in detail in \cite{perrard2021}.

\begin{figure}
\centering
\includegraphics[]{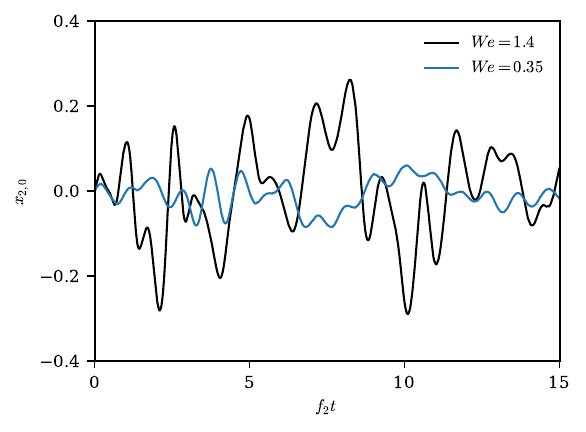}
\caption{Typical temporal evolution for the mode $(2, 0)$ at two different Weber numbers. Time is made dimensionless using the Rayleigh frequency $f_2$. Modes exhibit random oscillations, with an amplitude increasing with $\We$.}
\label{fig:ex}
\end{figure}

Figure~\ref{fig:ex} shows two typical temporal evolution for the mode $(\ell, m) = (2, 0)$, at two different Weber numbers.
Time is made dimensionless using the Rayleigh frequency $f_2$.
For both We, we observe random oscillations around zero and the predominance of the bubble resonant frequency $f_2$.
The amplitude of the oscillations increases with We.

Since we do not prescribe any special orientation relative to the bubble shape, all modes with the same principal number $\ell$ are statistically equivalent.
Indeed, one can verify that a rotation of a mode can be expressed as a linear combination of all the other modes with the same principal number.
As a consequence, we omit $m$ in what follows.
For instance $x_\ell(t)$ represents a typical temporal evolution of one of the modes $\ell$.
In addition, assuming that $x_{\ell, m}$ are independent, the ensemble averaging operation $\langle \cdot \rangle$ is computed over different simulations and over the $m$ values for a given $\ell$.
\citet{roa2023} used a reference frame dynamically oriented with the bubble principal axis of deformations.
In practice, their reference frame maximizes the amplitude of mode $(2,0)$, such that the differential elongation can be studied as the invariance by rotation is broken.

\section{Determination of the reduced dynamics}
\subsection{Model: a stochastic linear oscillator}
\label{subsec:model}
Following \citet{risso1998}, we introduce a linear stochastic model to describe each mode dynamics,
\begin{equation}
    \ddot{x}_{\ell} + \Lambda_\ell(\We) \dot{x}_\ell + \Omega_\ell(\We)^2 x_\ell = \force_\ell(\We, t),
    \label{eq:model}
\end{equation}
where $\Lambda_\ell$ and $\Omega_\ell$ are the damping rate and natural frequency respectively and $\force_\ell$ is a random variable which models the turbulent forcing.
In this section, we aim at measuring $\Lambda_\ell$, $\Omega_\ell$ and the statistical properties of $\force_\ell$ from the deformation dynamics.
Both parameters $\Lambda_\ell$ and $\Omega_\ell$, as well as the forcing statistics, may depend on the Weber number We.
Conversely to what other authors have done, time is made dimensionless using the eddy turnover time at the bubble scale $t_c(d) = \epsilon^{-1/3}d^{2/3}$ and, from now on, $\dot \cdot$ denotes derivatives with respect to this dimensionless time.
This choice avoid \textit{a priori} to have a forcing term depending on bubble properties such as surface tension.
It decorrelates the turbulent forcing (righ hand side), from the bubble response (the left hand side).
In these units, the Rayleigh frequency and the Lamb damping rate write $\omega_\ell^2 = 16(\ell -1)(\ell + 1)(\ell +2)/\We$ and $\lambda_\ell = 8 \sqrt{2}(\ell + 2)(2 \ell +1) \Re(d)^{-1}$ respectively.
Note that, in this study, we have not varied the eddy turnover time. 
When the bubble size lies within the inertial range of the turbulent cascade its dynamics is primarily controlled by inertial effects, and the parameters may not depend explicitly on the bubble Reynolds number, as long as $\Re(d) \gg 1$.

In order to measure the coefficients and the force statistics of equation~\eqref{eq:model}, we make the following assumptions:
\begin{description}
    \item[(H1): ]Modes dynamics are linear and uncoupled, which is valid for $x_\ell \ll 1$, corresponding to $\We\ll 1$.
    \item[(H2): ]The bubble deformation is one way-coupled to the flow. This hypothesis is discussed in section~\ref{subsec:forcing}.
    \item[(H3): ]The forcing $\force_\ell$ is statistically stationary.
    \item[(H4): ]The damping rate and the natural frequency do not depend on time.
\end{description}
From hypothesis (H2) the effective $\force_\ell$ in unit of the eddy turnover time, is independent of $\We$.
From hypothesis (H3), the effective forcing is completely determined by its auto-correlation function (or equivalently its spectrum), and its probability distribution function (pdf).\\

Under these hypothesis, in the next sections, we will show that
\begin{enumerate}
    \item The natural frequency is not modified by the presence of the flow: $\Omega_\ell = \omega_\ell$\label{result:omega}.
    \item There is an effective viscosity, driven by turbulence, so that $\Lambda_\ell =0.6 (\ell + 2)(2\ell + 1)$ for $\Re(d)=124$.
    \label{result:damping}
\end{enumerate}
Combining (\ref{result:omega}), (\ref{result:damping}) and equation~\eqref{eq:model} we will deduce the statistical properties of the forcing $\force_\ell$.

\subsection{Frequency response of the oscillator - Amplitude of the Fourier transform}

To rationalize the qualitative observations of figure~\ref{fig:ex} and identify the angular frequency $\Omega_\ell$, we investigate the frequency response of the bubble.
To do so, we compute the temporal Fourier transform, $\hat x_\ell$ of $x_\ell$ for all $\ell$ and $\We$,
\begin{equation}
    \hat{x}_\ell(\omega) = \int_{-\infty}^\infty x(t)e^{-i\omega t} \dd t,
\end{equation}
where $\hat x_\ell$ is also a random variable. Similarly, we  introduce $\hat \force_\ell$, the Fourier transform of the effective forcing,
\begin{equation}
    \hat{\force}_\ell(\omega) = \int_{-\infty}^\infty \force(t)e^{-i\omega t} \dd t
\end{equation}

\begin{figure}
\centering
\includegraphics[scale=1]{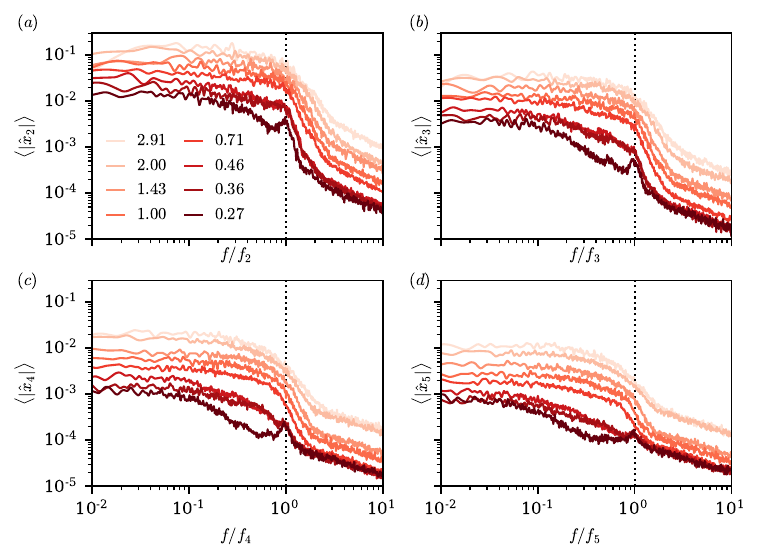}
\caption{Amplitude of the modes' Fourier transform for all We as a function of the frequency normalized by the corresponding Rayleigh frequency. The Weber number value is color-coded.}
\label{fig:fft}
\end{figure}

Figures~\ref{fig:fft} shows the ensemble average $\langle |\hat x_\ell| \rangle$ as a function of the frequency, $f = \omega/(2 \pi)$, normalized by the corresponding Rayleigh frequency, $f_\ell$. 

For $f<f_\ell$, for every $\We$, $\langle |\hat x_\ell| \rangle$ is approximately constant.
The low frequency dynamics is similar to that of a white noise.

At $f=f_\ell$ (back dotted line), for $\We\leq 0.46$ we observe a peak that resembles the resonant response of an oscillator at its natural frequency. This peak does not exist for larger values of We. Nevertheless, for every $\ell$, we observe a transition at this very frequency.

For $f>f_\ell$, at all We, we report a sharp power-law decay, following at least $(f/f_\ell)^{-4}$.

Finally, for $f> 3f_\ell$, the spectrum amplitude is above the noise level.
Note that this part also corresponds to the end of the inertial range.

Dimensional measurements of bubble deformation dynamics were performed by \citet{ravelet2011} in the context of bubbles rising in turbulence. They measured the temporal spectrum of the horizontal bubble main axis, a proxy for the amplitude of the second Rayleigh mode. The overall shape of their power spectrum was similar : weak variation for $f<f_2$, no resonance at $f_2$ and an a strong decay for $f>f_2$. In the absence of gravity, \citet{risso1998} also reported a transition at $f_2$, with a rapid decay for $f>f_2$ of the projected area spectrum.

The cut-off frequency being $f_\ell$ for all considered cases, we deduce that the bubble natural frequency in turbulence, $\Omega_\ell$ of equation~\eqref{eq:model}, is not modified by the presence of the surrounding turbulent flow and that,
\begin{equation}
    \Omega_\ell = \omega_\ell = 4 \left [\frac{(\ell -1)(\ell + 1)(\ell +2)}{\We}\right ]^{1/2}
    \label{eq:natfreq}
\end{equation}
It is surprising that the bubble natural frequency remains unchanged. Indeed, \citet{prosperetti1980} showed, for a bubble in an initially quiescent flow, that viscous effects induces an additional memory term in bubble dynamics.
This memory term can be modeled by an effective natural frequency and damping term.
The surrounding flow field can also modify the natural frequency.
In a uniaxial straining flow for instance, \citet{kang1988} demonstrated that a coupling between modes $\ell=2$ and $\ell=4$ decreases the mode 2 natural frequency at linear order, with a corrective term linear in We.
We hypothesize that the stochastic nature of turbulence cancels, in average, these contributions. In the following, we use the theoretical expression of $\omega_\ell$, for the bubble natural frequency, $\Omega_\ell$.

\subsection{Zero frequency limit and We-dependency of the forcing}

In this section, we investigate the zero frequency limit, and discuss the consequence for the We-dependency of the forcing.
By computing the Fourier transform of equation~\eqref{eq:model}, combined with \eqref{eq:natfreq}, we obtain an expression linking $\hat x_\ell$ and $\hat \force_\ell$,
\begin{equation}
    |\hat{x}_\ell|(\We, \omega) = \frac{|\hat{\force}_\ell|(\We, \omega)}{\sqrt{(\omega^2 - \omega_\ell(\We)^2)^2 + \Lambda_\ell(\We)^2 \omega^2}}
    \label{eq:fftmodel}
\end{equation}
The spectral behavior of each $x_\ell$ is a combination of the forcing spectrum $\hat \force_\ell$ and the bubble response.
In the limit case  $\omega=0$, using the expression of the bubble natural frequency~\eqref{eq:natfreq}, we have
\begin{equation}
    |\hat{x}_\ell|(\We, 0) = \frac{|\hat{\force}_\ell|(\We, 0)}{\omega_\ell(\We)^2}= \frac{\We}{16(\ell -1)(\ell + 1)(\ell +2)}  |\hat{\force}_\ell|(\We, 0).
    \label{eq:zerolimit}
\end{equation}
We can use this expression to investigate the We-dependency and $\ell$-dependency of $\hat \force_\ell$ at $\omega=0$. 
We extract $\langle|\hat{x}_\ell|\rangle(\We, 0) $ by averaging $\langle|\hat{x}_\ell| \rangle(\We, \omega)$ over the range $5.10^{-3}<f/f_\ell<10^{-1}$ where the spectrum is constant.

\begin{figure}
\centering
\includegraphics[]{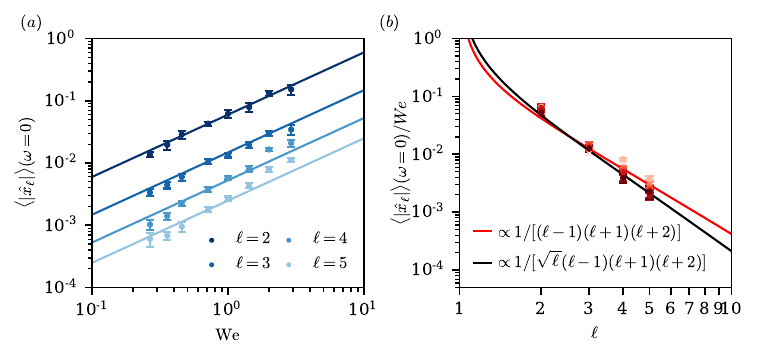}
\caption{a) Zero frequency limit of the modes' Fourier transform as a function of $\We$ for all $\ell$. Theoretical prediction $\avg{|\hat x_\ell|} (\omega=0) \propto \We$ is superimposed in solid lines. Error bars are estimated using the standard deviation of the spectrum value for $5.10^{-3}<f/f_\ell<10^{-1}$. b)~Compensated limit $\langle | \hat x_\ell| \rangle(\omega=0)/ \We$, as a function of $\ell$. Colors encode the We (see figure~\ref{fig:fft}). Assuming $\force_\ell$ independent of $\ell$ gives the scaling plotted in red.
Assuming $|\force_\ell| \sim 1/\sqrt{\ell}$ gives the scaling plotted in black.}
\label{fig:fftw0}
\end{figure}

Figure~\ref{fig:fftw0}a shows $\langle |\hat{x}_\ell|\rangle(\We, 0)$ as a function of $\We$. Solid lines of slope 1 are superimposed, showing that $\langle |\hat{x}_\ell|\rangle(\We, 0)$ increases linearly with $\We$ at all $\ell$. It follows from equation~\eqref{eq:zerolimit} that $\langle|\hat \force_\ell|\rangle (\We, 0)$ is independent of $\We$ for all $\ell$.
This result justifies that the effective forcing  from turbulence does not depend on bubble deformability at low frequency.
The modification of the flow induced by bubble oscillations does not impinge back on bubble dynamics.
A similar phenomenon has been observed for drops by \citet{vela2021}. They investigated the interfacial stress generated by eddies depending on their distance to the interface.
They concluded that eddies further that $0.2d$ from the drop interface (outer eddies) generate most of the stress.
They reported that these contributions are, in addition, independent of We, as these eddies are two far from the interface to be affected by drop deformations. We can assume that a similar mechanism may hold also for bubble dynamics so that $\hat \force_\ell$ does not depend on $\We$ either.
These results justify hypothesis (H2) at low frequency, and we assume that (H2) holds for all frequencies. From now on, we therefore assume that $\force_\ell$ does not depend on We. This hypothesis will be further validated and tested in section~\ref{subsec:forcing}.
The zero frequency limit also depends on the mode order $\ell$.  Figure~\ref{fig:fftw0}b shows the compensated spectrum limit $\langle | \hat x_\ell| \rangle(\omega=0)/ \We$ as a function of $\ell$.
We find that the zero frequency limit decreases slightly faster than $\omega_\ell^{-2} \sim [(\ell -1)(\ell +1)(\ell+2)]^{-1}$ of equation~\eqref{eq:zerolimit} (red line).
It suggests that $|\hat \force_\ell|$ weakly decreases with $\ell$, with $|\hat \force_\ell| \sim 1/\sqrt{\ell}$. Higher order modes are associated to smaller scales that are less energetic. However, the direct investigation of pressure modes in  section~\ref{subsec:pressuremodes} showed a much faster decreases of the mode energy with $\ell$. The high order modes $\ell \geq 3$ may also be indirectly forced from non linear coupling with the mode 2 changing the $\ell$-dependency of the forcing.

\subsection{Determination of the effective damping factor: Additional dissipation due to turbulence}
\label{subsec:zerofreqlim}
In this section, we present a method to compute the damping factor $\Lambda_\ell$ of equation~\eqref{eq:model} from the numerical data.

Let $\hat x_a$ and $\hat x_b$ be the Fourier transform $\hat x_\ell$ of the same mode $\ell$ for two Weber numbers $\We_a$ and $\We_b$.
For simplicity here, we denote $\omega_a$ and $\lambda_a$, the natural frequency and damping rate associated to $\We_a$ at this $\ell$.
Under hypothesis (H2), the ratio $R_{ab}$
\begin{equation}
    R_{ab}(\omega) = \left (\frac{\langle|\hat x_a|\rangle}{\langle|\hat x_b|\rangle} \right )^2 = \frac{(\omega^2 - \omega_b^2)^2 + \Lambda_b^2 \omega^2}{(\omega^2 - \omega_a^2)^2 + \Lambda_a^2 \omega^2}
    \label{eq:fftratio}
\end{equation}
is independent on $\hat \force_\ell$.

Since the two natural frequencies $\omega_a$ and $\omega_b$ are known (equation~\eqref{eq:natfreq}), one can estimate the two damping factors, $\Lambda_a$ and $\Lambda_b$, using $R_{ab}(\omega_a)$ and $R_{ab}(\omega_b)$, the ratios at the two natural frequencies
\begin{align}
    R_{ab}(\omega_a) & = \frac{(\omega_a^2 - \omega_b^2)^2 + \Lambda_b^2 \omega_a^2}{ \Lambda_a^2 \omega_a^2} \label{eq:RatiosA}\\
    R_{ab}(\omega_b) & = \frac{\Lambda_b^2 \omega_b^2}{(\omega_a^2 - \omega_b^2)^2 + \Lambda_a^2 \omega_b^2}
    \label{eq:RatiosB}
\end{align}
by solving this two-equations system. Note that an optimization of $\Lambda_a$ and $\Lambda_b$ on the whole range of frequencies was less reliable. The signal over noise ratio is optimal near the resonance, and decreases both at high and low frequencies. Indeed, high frequencies, which are the more noisy, then dominate the optimization procedure. 

\begin{figure}
\centering
\includegraphics[]{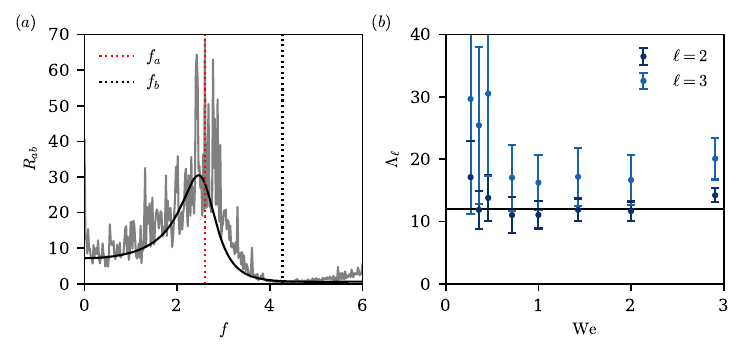}
\caption{a)~Ratio between the Fourier spectrum at $\We_a = 0.71$ and $\We_b = 0.27$ for the mode $\ell=2$. The red and black vertical lines denotes the position of the Rayleigh frequency at these two We where we evaluate $R_{ab}$. The black line is the prediction from equation~\eqref{eq:fftratio}. b)~Damping factor as a function of We for $\ell=2$ and $\ell =3$. The error bars represent the standard deviations. The solid black line corresponds to $\Lambda_2 = 12$.}
\label{fig:damping}
\end{figure}

Figure~\ref{fig:damping}a illustrates the computation of $\Lambda_\ell$. The ratio $R_{0.71, 0.27}$ for $\ell = 2$, $\We_a = 0.71$ and $\We_b = 0.27$ is represented as a function of the frequency $f$ (grey curve). The black and red vertical lines denote the position of the two natural frequencies $\omega_a$ and $\omega_b$ respectively, at which we measure $R_{0.71, 0.27}$.
Inverting system \eqref{eq:RatiosA}-\eqref{eq:RatiosB} gives an estimate of $\Lambda_{0.71}$ and $\Lambda_{0.27}$.
Using these computed values of $\Lambda_{0.71}$ and $\Lambda_{0.27}$ we plot the theoretical expression of equation~\eqref{eq:fftratio} at all frequencies (black line).
This expression captures the main features of the ratio $R_{0.71, 0.27}(\omega)$: the low frequency limit, the position and amplitude of the peak.

We then follow this procedure for every pair $(\We_a, \We_b)$ and obtain 14 estimations of $\Lambda_\ell$ per Weber number per mode $\ell$. We did not find a significant bias on the estimated value of $\Lambda_\ell(\We)$ as a function of the Weber ratio $\We_a/\We_b$.
We then average over all values of $\We_b$ values to estimate $\Lambda_\ell(\We_a)$. The values of $\Lambda_\ell$ for $\ell=2$ and $\ell=3$ as a function of $\We$, and their standard deviation are reported in table~\ref{tab:ValLambda}.
\begin{table}
    \centering
    \begin{tabular}{|c | c c c c c c c c |} 
    \hline
     We & 2.9 & 2 & 1.43 & 1 & 0.71 & 0.46 & 0.36 & 0.27 \\ [0.5ex] 
     \hline
     $\Lambda_2$ & 14.2 & 11.7 & 11.9 & 11.1 & 11.0 & 13.8 & 11.9 & 17.1\\ [0.5ex]
     \hline
     $\sigma_\Lambda^2$ & 1.1 & 1.6 & 1.8 & 2.2 & 2.9 & 3.6 & 3.1 & 5.9 \\[0.5ex]
     \hline
     $\Lambda_3$ & 20.1 & 16.7 & 17.2 & 16.2 & 17.0 & 30.5 & 25.4 & 29.7\\ [0.5ex]
     \hline
     $\sigma_\Lambda^3$ & 3.4 & 4.0 & 4.6 & 4.4 & 5.3 & 13.2 & 12.6 & 18.4 \\[0.5ex]
     \hline
     \end{tabular}   
     \caption{Average damping parameter $\Lambda_\ell$ and corresponding standard deviation, for every We.}
    \label{tab:ValLambda}
\end{table}
For $\ell\geq 4$, equation~\eqref{eq:fftmodel} fails to describe the ratio $R_{ab}$. Figure~\ref{fig:damping}b shows $\Lambda_\ell$ as a function of We for $\ell=2$ and $\ell=3$ with errorbars encoding the standard deviation $\sigma_\Lambda^\ell$. We find no clear variation of $\Lambda_\ell$ with We, especially for $\ell=2$.
When $\ell$ increases, the dissipation also increases, as smaller scales are more efficient to dissipate energy. The increase of $\Lambda_\ell$ with $\ell$ is compatible with the $\ell$-dependency in a quiescent environment from \citet{lamb1932}. 
From our observations we found the following expression for the damping factor,
\begin{equation}
    \Lambda_\ell =0.6 (\ell + 2)(2\ell + 1).
    \label{eq:damping}
\end{equation}
In quiescent environments, dissipation is also independent on We, $\lambda_\ell =8 \sqrt{2}(\ell + 2)(2 \ell +1) \Re(d)^{-1}$, as it originates from molecular diffusion in the liquid.
However, we find $\Lambda_2 \approx 13 \lambda_2$.
The surrounding flow field induces an additional effective damping.
Experimentally, \citet{ravelet2011} also observed an additional dissipation for bubbles rising in turbulence but attributed it to the presence of the wake.
Yet, similar observations come from drop oscillations in space.
In the presence of a turbulent internal flow, drop oscillations are significantly damped
\citep{bojarevics2003,berry2005}.
This additional dissipation is interpreted in terms of a turbulent eddy viscosity \citep{xiao2021}.
In addition, \citet{vela2021} showed that there is a transfer of energy from the drop interface to eddies closer than $0.2 \, d$ from the drop interface and inside the drop.
They call them inner eddies.
These small eddies efficiently dissipate energy.
This transfer of energy suggests that the enhanced dissipation comes from an increase in the local effective diffusivity.

It is advantageous to estimate the size of an equivalent mixing length $L_t$. This characteristic length of momentum transport has first been introduced by Prandtl~\citep{boussinesq1877,prandtl1949,xiao2021} to describe, in a turbulent flow, the logarithmic profile of velocity near a wall.
By dimensional considerations, one can estimate the effective turbulent viscosity $\nu_t$, using $L_t$ and a typical velocity scale of velocity fluctuations at that scale, $\langle \delta u(L_t)^2 \rangle^{1/2} $,
\begin{equation}
    \nu_t = \langle \delta u(L_t)^2 \rangle^{1/2} L_t = \sqrt{2}\epsilon^{1/3}L_t^{4/3}.
\end{equation}
Expressing the effective damping rate in terms of this effective dissipation gives,
\begin{equation}
    \Lambda_\ell = 8(\ell + 2)(2\ell +1)\frac{\nu_t d^{2/3}}{d^2 \epsilon^{1/3}} = 8 \sqrt{2}(\ell + 2)(2\ell +1) \bigg[\frac{L_t}{d}\bigg ]^{4/3}.
\end{equation}
Injecting equation~\eqref{eq:damping}, gives an estimate for $L_t$,
\begin{equation}
    L_t = \frac{d}{10} = \frac{\R}{5}.
    \label{eq:Ldiss}
\end{equation}
Being of the same order of magnitude as the bubble radius, we hypothesize that the mixing length originates from a geometric effect, similar to the separation between inner and outer eddies from \citet{vela2021}. We further investigate the origin of this dissipation in the last section, by looking at the local velocity gradients close to the bubble interface.

\subsection{Effective forcing statistics: Temporal correlations and distribution}
\label{subsec:forcing}

Since the left hand side of equation~\eqref{eq:model} is now completely determined, we can compute the right hand side, and interpret it as a forcing term from the turbulent flow.

For a bubble of size lying within the inertial range, the bubble deformations are known to be primarily driven by Eulerian pressure gradients integrated over the bubble surface \citep{qureshi2007}. In the limit of small bubble deformations, the pressure field at the bubble surface can be approximated by the pressure field on a sphere of radius $\R$.
Decomposed in the spherical harmonics base, this pressure reads
\begin{equation}
    p(\theta, \phi) = P_c \big[\sum_{\ell = 0}^\infty \sum_{m = -\ell}^\ell P_{\ell, m}(t)\,Y_\ell^m(\theta, \phi)\big ],
\end{equation}
where $P_c = \rho \delta u (d)^2$.
There is no direct experimental measurement of these pressure coefficients.
Nevertheless, for the modes $\ell = 1$, one can circumvent this difficulty by looking at the acceleration statistics of finite-size particles in a turbulent flow. Indeed, in the inertial range, Lagrangian acceleration of particles reflects the statistics of the force driving them, namely, the pressure gradient averaged over the particle volume \citep{qureshi2007,calzavarini2009}. This hydrodynamic force turns out to be the surface averaged pressure, which drives translational bubble motions.
For the higher order modes ($\ell\geq 2$), which drive bubble deformations, there is no equivalent measurements in turbulence.
Therefore, from time to time, we will compare our statistics of $\force_\ell$ ($\ell \geq 2$) with statistical quantities closely related to $P_1$, namely the Lagrangian acceleration statistics and the pressure increments.
A direct measure of the statistics of $P_\ell$ ($\ell\geq 2$) in the absence of bubble is provided in section~\ref{section:flow}.

Practically, we compute $\force_\ell$ from the modes' Fourier transform $\hat x_\ell$ using the following relation
\begin{equation}
    \force_\ell(t) = \frac{1}{2\pi}\int_{-\infty}^\infty \hat{x}_\ell(\omega)(\Omega_\ell^2 - \omega^2 + i \Lambda_\ell \omega) e^{i \omega t} \dd \omega,
\end{equation}
where we use the expressions of $\Lambda_\ell$ and $\Omega_\ell$ from Eqs. \eqref{eq:natfreq} and \eqref{eq:damping}.

\begin{figure}
\centering
\includegraphics[]{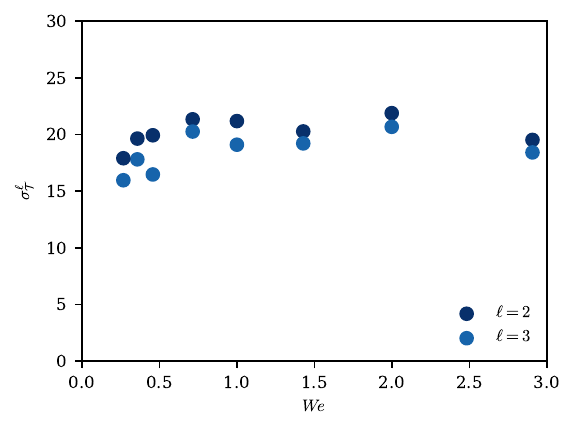}
\caption{Standard deviation of $\force$ as a function of We for $\ell=2$ and $\ell =3$. No We-dependency is observed, while $\sigma_\force$ decreases slightly for larger $\ell$.}
\label{fig:stdforce}
\end{figure}

As expected from rotational invariance, we find that the average forcing $\langle \force_\ell \rangle$ vanishes for all $\We$.
The standard deviation of $\force_\ell$, $\sigma_\force^\ell$ is shown in figure~\ref{fig:stdforce} as a function of $\We$ for $\ell =2$ and 3 (color-coded). $\sigma_\force^\ell$ is found to be almost independent of the Weber number.
We found that the effective forcing from the turbulent flow does not depend on bubble deformability. Therefore, bubble deformations are only one-way coupled to the flow.

In physical units, the force $\force_\ell$ then scales as $\alpha(\ell) \epsilon^{2/3}d^{-1/3}$, where $\alpha_\ell$ is a function of the mode order. The standard deviation slightly decreases with $\ell$, compatible with $\alpha_\ell \sim {\ell}^{-1/2}$.

In the context of Lagrangian particle acceleration in turbulence, the standard deviation of acceleration also decreases with particle size as $d^{-1/3}$. 
This scaling can be predicted using a scale invariant pressure fluctuations argument \citep{voth2002,qureshi2007,volk2011}.
In addition, Lagrangian acceleration statistics do not depend explicitly on the Reynolds number at the particle size $\Re(d)$, as long as the particle lies within the inertial range.
Only a marginal effect of the flow Taylor Reynolds number $\Re_\lambda$ on the variance of the acceleration \citep{voth2002} was found. As a consequence, we expect the effective forcing to be independent of $\Re_\lambda$, $\Re(d)$ and the Weber number.

\begin{figure}
\centering
\includegraphics{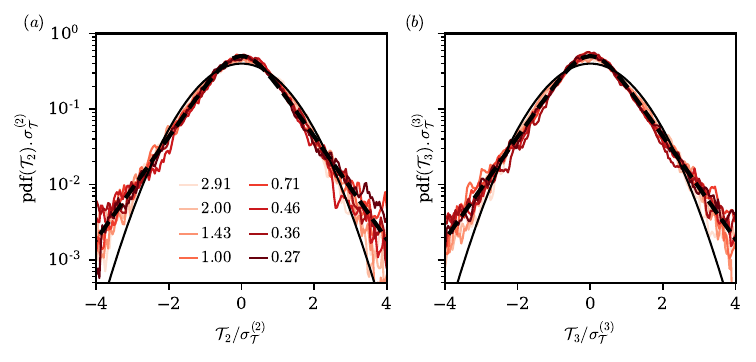}
\caption{Probability density functions of $\force_2$ (a) and $\force_3$ (b) for all We (color-coded), normalized by their standard deviations. The shape of the distribution is independent of We. The dashed line represents the hyperbolic secant distribution, while the solid line is the Gaussian distribution.}
\label{fig:pdfforce}
\end{figure}

Beyond the first two moments of the effective forcing distribution, it is interesting to look at the full distribution. 
Figures~\ref{fig:pdfforce}a and \ref{fig:pdfforce}b show the probability distribution of $\force_2$ and $\force_3$ respectively for all We, normalized by their standard deviation $\sigma_\force^\ell$.
We find that the shape of the distribution is also independent of the Weber number.
These distributions are characterized by exponential tails, and are well described by the hyperbolic secant distribution (black dashed line),
\begin{equation}
    \text{pdf}(\force) = \frac{1}{2\sigma_\force^\ell}\sech(\frac{\pi}{2}\frac{\force}{\sigma_\force^\ell})
    \label{eq:distribthforce}
\end{equation}
which depends on a single parameter, the standard deviation $\sigma_\force^\ell$.
The probability that a large forcing occurs is way larger than that of a Gaussian distribution (solid black line).

It is again tantalizing to compare this distribution to Lagrangian acceleration statistics for both particles and bubbles~\citep{voth2002,qureshi2007,volk2008,homann2010,prakash2012,salibindla2021}. For small, neutral tracers and particles of Kolmogorov scale size, the acceleration distributions exhibit larger tails, decreasing slower than exponential. However, for larger particles ($d/\eta > 10$), the shape exhibits exponential tail, independent of bubble size and therefore of $\Re(d)$ \citep{voth2002,qureshi2007,volk2011}.
The pdf shape of the Lagrangian acceleration is well described by the following expression, initially proposed for tracer particles \citep{mordant2004,qureshi2007}
\begin{equation}
    \text{pdf}(x) = \frac{e^{3s^2/2}}{4\sqrt{3}} \left [ 1 - \erf \big(\frac{\log(|x/\sqrt{3}|) + 2s^2}{\sqrt{2}x} \big)\right ]
    \label{eq:accpressure}
\end{equation}
where $x$ is the standardized variable and $s$ an additional fitting parameter. In the range of resolved scale, the two expressions, equations~\eqref{eq:distribthforce} and \eqref{eq:accpressure}, are compatible with our experimental data.

\begin{figure}
\centering
\includegraphics[]{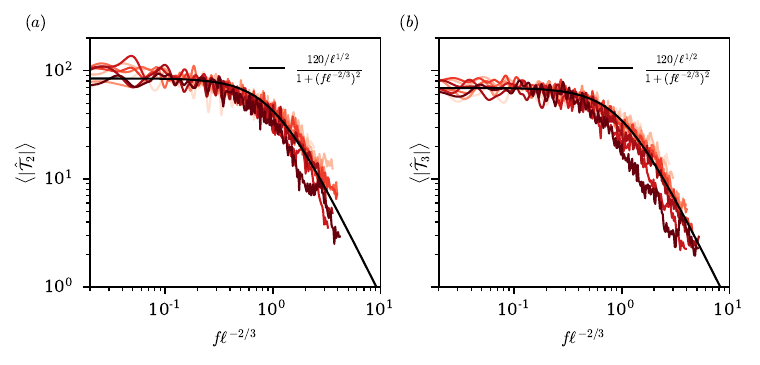}
\caption{Effective forcing spectrum for $\ell=2$ (a) and $\ell = 3$ (b) deduced from equation~\eqref{eq:deduceforce} as a function of the frequency normalized by the eddy turnover time at scale $d/\ell$. The Weber number is color-coded with the same colorbar as in figure~\ref{fig:fft}).}
\label{fig:forcespectrum}
\end{figure}

To characterize the temporal evolution of the effective forcing $\force_\ell$, we study its ensemble averaged Fourier transform $\langle |\hat \force_\ell| \rangle$. Injecting equations~\eqref{eq:natfreq} and \eqref{eq:damping} within equation~\eqref{eq:fftmodel} we obtain an expression in Fourier space for $\langle |\hat \force_\ell| \rangle$:
\begin{equation}
    \langle |\hat \force_\ell| \rangle = \langle |\hat x_\ell| \rangle. \left [{(\omega^2 - \Omega_\ell(\We)^2)^2 + \Lambda_\ell^2 \omega^2} \right]^{1/2}.
    \label{eq:deduceforce}
\end{equation}
Figure~\ref{fig:forcespectrum}a and \ref{fig:forcespectrum}b show $\langle |\hat \force_2| \rangle$ and $\langle |\hat \force_3| \rangle$ respectively as a function of $f \ell^{-2/3}$, where $\ell^{2/3}$ is the eddy turnover time at scale $d/\ell$ (in units of $t_c(d)$).
For all frequencies, we found that the effective forcing spectrum does not depend on the Weber number.
At low frequencies ($f<0.2 \, \ell^{2/3}$), the forcing amplitude is constant, corresponding to a white noise. For $f>\ell^{2/3}$, the decay of $\langle |\hat \force_\ell| \rangle$ is compatible with $1/f^2$. 
The limit between these two regimes is set by the eddy turnover time at scale $d/\ell$. We found that the spectrum of the effective forcing only depends on the turbulence parameters, and is therefore independent of the bubble deformations.
As was anticipated in section~\ref{subsec:model}, model~\eqref{eq:model} decouples the turbulent forcing (the right hand side) from the bubble response (the left hand side). The observation of a cut off frequency at the characteristic time scale of turbulent fluctuations at the mode scale $d/\ell$ can be interpreted as a filtering process originating from the integration over the bubble surface. This filtering operation is further discussed in section \ref{section:flow}.

From the previous observations, we propose the following expression for the forcing spectrum, 
\begin{equation}
    \langle |\hat \force_\ell| \rangle(f) = \frac{\tau_\ell}{ 1 + [f \ell^{-2/3}]^2},
    \label{eq:modelfftforce}
\end{equation}
where $\tau_\ell$ is a numerical constant, accounting for the $\ell$-dependency of $\force_\ell$, that is adjusted on the data. From equation~\ref{eq:zerolimit} and figure~\ref{fig:fftw0}, we estimate $\tau_\ell \sim \ell^{-1/2}$.
The expression~\eqref{eq:modelfftforce} captures quantitatively the effective forcing spectrum (solid black line in figures~\ref{fig:forcespectrum}a and \ref{fig:forcespectrum}b).

In the context of Lagrangian particle accelerations, \citet{voth2002} followed by \citet{volk2008}, computed the temporal autocorrelation of inertial particle accelerations in turbulence. The temporal acceleration statistics of a finite size particle is usually attributed to a filtering effect of the small scale turbulent fluctuations at the particle scale~\citep{qureshi2007}.
As a consequence, the correlation time of acceleration for neutrally buoyant particle is given by the eddy turnover time $t_c(d)$. This result has been recently extended to buoyant particle that exhibits a modified correlation time $t \sim t_c(d) \beta^{-1/2}$ \citep{fan2024}, where $\beta = 3 \rho/(2 \rho + \rho_p)$ is a function of the fluid density $\rho$ and the particle density $\rho_p$. For a bubble, we have $\beta = 3$, corresponding to a correlation time of order $t_c$. 
In our case, the temporal auto-correlation function $C_{\force_\ell}(t) = \langle \force_\ell(0) \force_\ell(t)\rangle/ (\sigma_\force^\ell)^2$  for the modes $\ell>1$ can be deduced from the spectrum $\hat \force_\ell$ and is written as
\begin{equation}
C_{\force_\ell}(t) = \exp(-2\pi \ell^{2/3}t)(1 + 2\pi \ell^{2/3}t).
\label{eq:autocorrforce}
\end{equation}
We found that the correlation time in physical units is given by $t_c(d) \ell^{-2/3}/(2 \pi)$, which also scales as $t_c(d)$, with an additional dependency in the mode order $\ell$. The prefactor being smaller than one, the mode oscillations decorrelate faster that the velocity fluctuations at the bubble scale. 

In summary, we found that all the statistics of $\force_\ell$ are independent of $\We$ which confirms the initial intuition of \citet{risso1998} that bubble dynamics and turbulent forcing are decoupled. We found that the bubble deformation by the flow field can be described by a one-way coupling model: the flow field generated by bubble oscillations does not significantly impinges back on bubble dynamics.
In addition, experimental results from the literature suggest that these statistics are likely to be independent on $\Re(d)$, as long as we consider bubbles larger than the Kolmogorov scale.

From the stationary hypothesis \textrm{(H3)}, the forcing is completely characterized by its distribution and temporal autocorrelation function. The combination of an explicit form for the pdf (eq.~\eqref{eq:distribthforce}) and for the autocorrelation function (eq.~\eqref{eq:autocorrforce}) then provides a complete model of a synthetic stochastic effective forcing for bubbles deformations in turbulence. Previous modelling approaches have used two points velocity measurements to model an effective forcing term~\citep{risso1998,lalanne2019,masuk2021simultaneous}, following the original idea from \citet{kolmogorov1949} and \citet{hinze1955}.
Here we found that the statistics of the effective forcing differ significantly from two points statistics, in particular due to the volumetric filtering effect at the particle size.

\section{Model validation}

To describe the bubble deformation, we have inferred step by step an equation including, damping, natural frequency and a statistical model for the effective forcing term $\force_\ell$. To validate our model, we check against the DNS data the predictions of our linear model.

\subsection{Modes' standard deviation and distributions}

\begin{figure}
\centering
\includegraphics[]{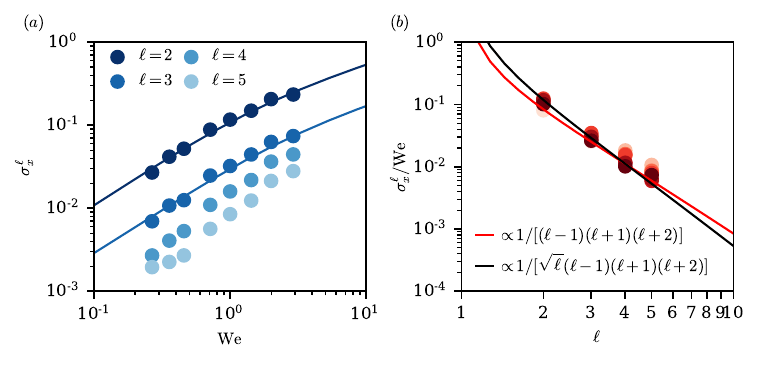}
\caption{a) Modes standard deviation as a function of the Weber number for all $\ell$. The two straights lines are the predictions from our linear model, for modes $\ell=2$ and $\ell=3$. b)~Modes' standard deviation compensated by We as a function of the mode principal number.}
\label{fig:std}
\end{figure}

We first look at the modes' standard deviation $\sigma_x^\ell$ and statistics. Figure~\ref{fig:std}a shows the modes's standard deviation as a function of the Weber number for $\ell \in [2, 5]$. We find that $\sigma_x^\ell$ can be approximated by $\sigma_x^\ell \approx  \We/[(\ell-1)(\ell+1)(\ell+2)]$, with a constant of order one. The prediction from our linear model computed from an integration over the Fourier space, is superimposed in solid line for $\ell = 2$ and 3 using eq.~\eqref{eq:model}, showing a quantitative agreement with the numerical data. 

A scaling for $\sigma_x^\ell$ as a function of $\We$ and $\ell$ can be deduced analytically in model cases. One natural case is to consider $\force_\ell$ as a Gaussian white noise of autocorrelation function $C(t) = D \delta (t)$, where $\delta$ is the Dirac function, and $D$ is independent of the Weber number. In this case, from the analysis of stochastic harmonic oscillators \citep{gitterman2005} the standard deviation reads
\begin{equation}
\sigma_x^\ell \sim \left [ \frac{D}{\Lambda_\ell \Omega_\ell^2}\right ]^{1/2}.
\end{equation}
From the coefficients $\Lambda_\ell$ and $\Omega_\ell$ we extracted, this model predicts $\sigma_x^\ell \propto \We^{1/2}$, which does not correspond to the observed scaling. A finite correlation time has be taken into account. We then consider $\force_\ell$ as an exponentially correlated Gaussian noise of autocorrelation function $\langle \force_\ell(t) \force_\ell(t^\prime)\rangle = (\sigma_\force^\ell)^2 \exp( - |t - t^\prime|/t_\ell)$, where $t_\ell = \ell^{-2/3}/(2 \pi)$ is the correlation time of the effective forcing deduced from equation~\ref{eq:autocorrforce}, and $D$ is independent of $\We$. In this case the mode's standard deviation reads \citep{gitterman2005}, 
\begin{equation}
    \sigma_x^\ell = \sigma_\force^\ell \left[\frac{ t_\ell (1 + \Lambda_\ell t_\ell)}{ \Omega_\ell^2 \Lambda_\ell( 1 + \Lambda_\ell t_\ell + \Omega_\ell^2 t_\ell^2)} \right ]^{1/2}. 
    \label{eq:stdcolorednoise}
\end{equation}
The scaling of $\sigma_x^\ell$ now becomes a function of the ratios $\Lambda_\ell t_\ell$ and $\Omega_\ell t_\ell$. In practice, we have $\Omega_\ell t_\ell \gg 1$ and $\Omega_\ell t_\ell \gg \Lambda_\ell t_\ell$ for sufficiently small Weber. Considering the limit $\Lambda_\ell t_\ell \gg 1$, equation~\eqref{eq:stdcolorednoise} simplifies as
\begin{equation}
    \sigma_x^\ell =\frac{\sigma_\force^\ell}{\Omega_\ell^2} = \frac{\sigma_\force^\ell}{(\ell -1)(\ell + 1)(\ell +2)}\We.
\end{equation}
We then recover the observed scaling for small Weber number. For larger Weber number, the ratio $\Omega_\ell t_\ell$ decreases, and we expect a transition to a shallower increase of $\sigma_x^\ell$ with We. This transition should occur for larger Weber number as $\ell$ increases, an interpretation compatible with the numerical data shown in figure~\ref{fig:std}a. The observed scaling of $\sigma_x^\ell$ with Weber thus corresponds to a saturation of the bubble deformations dominated rather by the long correlation time of the forcing (frozen turbulence hypothesis applied to bubble deformations~\citep{ruth2019}) than an accumulation of random forcing events on a time scale $1/\Lambda_\ell$. It is worth noticing that the estimate of the correlation time of the forcing is therefore essential to predict the amplitude of bubble deformations.

To further check the dependency in $\ell$, figure~\ref{fig:std}b shows the compensated standard deviation $\sigma_x^\ell /\We$. We recover that the decrease of the modes' amplitude with $\ell$ can be mainly attributed to the increase of the natural frequency with $\ell$, with a small correction originating from the weak dependency of $\force_\ell$ with $\ell$. Eventually, we found a quantitative agreement between the standard deviation $x_\ell$ and the predicted value from the linear model. The model captures the evolution of $\sigma_x^\ell$ with both We and $\ell$.

The linear increase of $\sigma_x^\ell$ with We, up to the critical Weber number ($\We \approx 3$ in our case) has important consequences when modelling bubble breakup.
\citet{risso1998} suggested that the threshold for breakup is close to the value above which the deformations start to be non linear.
A linear model would then be sufficient to describe bubble deformations up to the breakup threshold.

\begin{figure}
\centering
\includegraphics{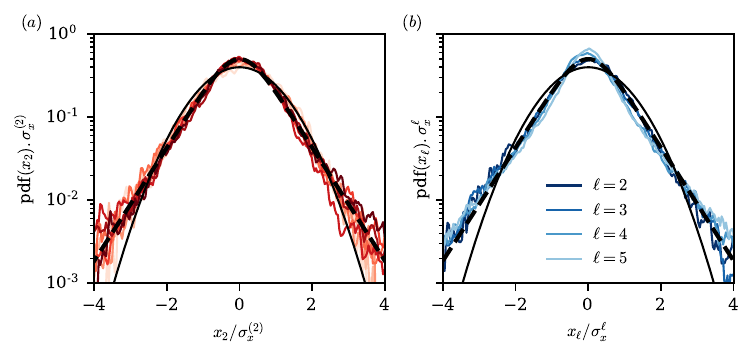}
\caption{a) Normalized pdf of $x_2$ for all We. b)~Normalized pdf at $\We=1$ for different $\ell$. a)\&b) The black dashed line is the hyperbolic secant distribution. The solid black line the Gaussian distribution.}
\label{fig:pdf}
\end{figure}

We then look at the entire statistics of the $x_\ell$. Figures~\ref{fig:pdf} show the probability density functions of the modes $\ell=2$ for all Weber numbers (\ref{fig:pdf}a) normalized by their standard deviation. We find that the shape of the pdf does not depend on the Weber number and corresponds to the hyperbolic secant distribution (black dashed line), equivalent to the pdf of the effective force $\force_\ell$. Both the forcing and the mode amplitude share the same pdf that deviate from gaussianity (solid black line) with exponential tails. As the distributions exhibit fat tails, the probability that bubbles experience large deformations leading to breakup is large compared to a Gaussian distribution (black dotted line). Moreover, for larger $\ell$, the deviation from gaussian distribution increases, as shown in figure \ref{fig:pdf}b for $\We=1$.

\subsection{Deformation spectrum}
\begin{figure}
\centering
\includegraphics{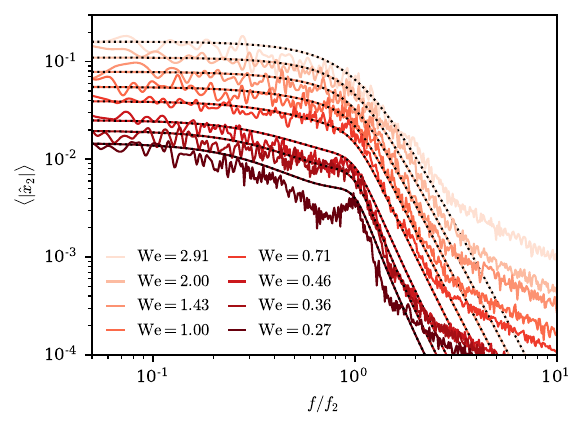}
\caption{Comparison of the Fourier spectrum amplitude between the DNS and the model (dotted line). The model spectrum is obtained by combining equations~\eqref{eq:damping}, \eqref{eq:natfreq} and \eqref{eq:modelfftforce} in \eqref{eq:fftmodel}, for the mode $\ell=2$. The model captures the low frequency limit, the position of the transition as well as the high frequency decay for all We.}
\label{fig:compspectrum}
\end{figure}

Figure~\ref{fig:compspectrum} compares the modes' Fourier transforms with the model prediction~\eqref{eq:fftmodel} combined with equations~\eqref{eq:damping}, \eqref{eq:natfreq} and \eqref{eq:modelfftforce} (dotted lines).
For all Weber numbers, the model accurately predicts the zero-limit frequency as well as the amplitude of the spectrum at the bubble natural frequency $f_2$ and the position and slope of the decay at larger frequencies. At the lowest Weber number ($\We = 0.27$), the model overestimates the spectrum just below the resonance. We remind here that for frequencies larger than $2.5f_2$ the spectrum is dominated by numerical noise.  For all the other We, in the absence of resonance, the model captures the spectrum close to the bubble natural frequency.

\subsection{Consequences for bubble breakup}
Thanks to the quantitative model we develop, we can revisit the breakup scenario and the criterion for breakup. Two main breakup scenarios have been proposed for bubbles in turbulence. Bubbles can break either when they encounter a pressure fluctuation larger than some threshold values \citep{lee1987,luo1996,wang2003,masuk2021model} or after series of small excitation at their natural frequency which induce a resonance \citep{sevik1973,risso1998}.
The ability for a bubble to store energy on a mode $\ell$, is quantified by the quality factor $Q_\ell = \Omega_\ell/\Lambda_\ell$. For large quality factor typically $Q_\ell>10$, modes can store energy over several periods of oscillation, while for lower quality factor, energy is dissipated in few bubble periods at most. Our  linear  model  provides a quantitative measure of $Q_\ell$.
Combining equations~\eqref{eq:natfreq} and \eqref{eq:damping} we have an explicit expression for $Q_\ell$ as a function of We and $\ell$,
\begin{equation}
    Q_\ell = 4\sqrt{\frac{(\ell - 1)(\ell +1)}{0.6(\ell +2)(2\ell +1)^2}}\We^{-1/2}.
\end{equation}
In turbulence, bubbles mainly break after oblate-prolate deformations, meaning deformations along their second modes $\ell=2$ \citep{risso1998,ravelet2011,perrard2021,masuk2021simultaneous}. For the typical critical Weber numbers reported in the literature, $0.1 <\We_c <10$ \citep{sevik1973,risso1998,martinez1999a,riviere2021}, our estimate of the quality factor for the mode $\ell=2$ ranges from $0.3$ ($\We_c = 10$) to 3 ($\We_c = 0.1$). These quality factors are too low to observe significant energy storage over several period of oscillations. We conclude that large pressure fluctuations set the value of the critical Weber number rather than resonant mechanism.

Note that a sequence of oscillations at the bubble natural frequency may be observed for sufficiently large quality factor, typically $Q_2>10$, corresponding to $\We<8.10^{-3}$. Even though such a Weber number corresponds to bubbles size much smaller than the Kolmogorov Hinze scale, which will never break, it may be observed experimentally.

\section{Link between model coefficients and surrounding turbulent fields}
\label{section:flow}

In this section, we aim at connecting the effective variables we identified, namely the forcing  $\force_\ell$ and the damping coefficient $\Lambda_\ell$, to relevant flow statistics in turbulence. For the effective force, bubble deformations are known to originate from pressure differences along the interface \citep{qureshi2007}. The presence of a bubble modifies the flow statistics in its surrounding, through dynamical boundary conditions at the interface and incompressibility. Nevertheless, for drops, it has been shown that the outer eddies (further than $0.2d$ from the interface) generate most of the normal stress \citep{vela2021}. These outer eddies may be less affected by the presence of the interface. We will therefore assume that the pressure statistics in the absence of bubble are a reasonable proxy to estimate $\force_\ell$. On the other hand, the dissipation is expected to originate from the boundary layer near the interface~\citep{vela2021}. To rationalise the origin of the additional damping  from the flow statistics, we will therefore study the local dissipation in the bubble boundary layer.

\subsection{Point statistics of the pressure field}

As a reference case, let us first consider the Eulerian point statistics of pressure in homogeneous and isotropic turbulence. 
To compare with the bubble dynamics, we will still express length scales in units of $d$, timescales in units of $t_c(d)$ and therefore velocity in term of velocity increments at the bubble scale $\avg{\delta u(d)^2}^{1/2}$.

We run single phase direct numerical simulations and record the Eulerian pressure fluctuations $p(x, t)$ at seven different fixed location well separated in space. We run three simulations for a total of $245 t_c(d)$. Resolution is increased compared to the two-phase problem and would be equivalent to 68 points per bubble radius and 3.6 points per Kolmogorov length. In this section, ensemble averages are performed over the three simulations and the seven locations.

\begin{figure}
\centering
\includegraphics[]{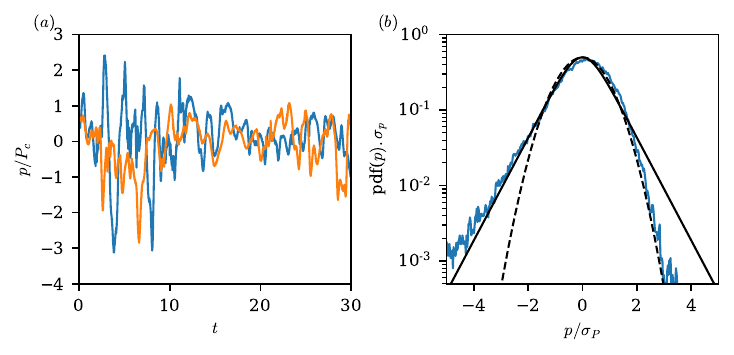}
\caption{a)~Typical temporal evolution of the pressure at two points in space. We observe small amplitude oscillations together with rare intense negative peaks. b)~Local pressure distribution normalized by its standard deviation $\sigma_p = 0.67 P_c$. The solid black line follows the hyperbolic secant distribution centered while the black dashed line follows a Gaussian distribution with standard deviation $4/5$.}
\label{fig:ploc-pdf}
\end{figure}

Figure~\ref{fig:ploc-pdf}a illustrates two temporal evolution of pressure, normalized by the characteristic pressure difference at the bubble scale, $P_c= \rho \delta u(d)^2$. We found a pressure standard deviation $\sigma_p = 0.67 P_c$. Pressure exhibits random oscillations of small amplitude around zero, together with large negative drops. This asymmetry between positive and negative pressure fluctuations is better observed on the pressure pdf plotted on figure~\ref{fig:ploc-pdf}b. We recover that negative values are exponentially distributed, while positive pressure values follow a Gaussian distribution (dashed black line).
The existence of the large negative peaks leading to an asymmetric pdf is well known and has been reported both in experiments \citep{abry1994,pumir1994,cadot1995} and direct numerical simulations of homogeneous isotropic turbulence \citep{cao1999,vedula1999}.
It has been shown that these large negative peaks correspond to vorticity filaments~\citep{douady1991,fauve1993,cadot1995} passing through the measurement point. As the bubble moves in the fluid, it may experience different pressure statistics and the Lagrangian pressure statistics could also be relevant.

Lagrangian pressure statistics have also been investigated numerically. Numerical studies involve measuring pressure statistics along the paths of point particles \citep{bappy2019}, as well as (sub-Kolmogorov) finite-size bubbles \citep{bappy2020effect,bappy2020pressure} whose dynamics are modeled using a pure advection or a Maxey-Riley equation \citep{maxey1983,toschi2009} respectively. They found that larger particles have a higher probability to be within low pressure regions. Nevertheless, the overall shape of the pressure pdf, with an exponential tail for negative values and a Gaussian distribution of positive values, is conserved.

\begin{figure}
\centering
\includegraphics[]{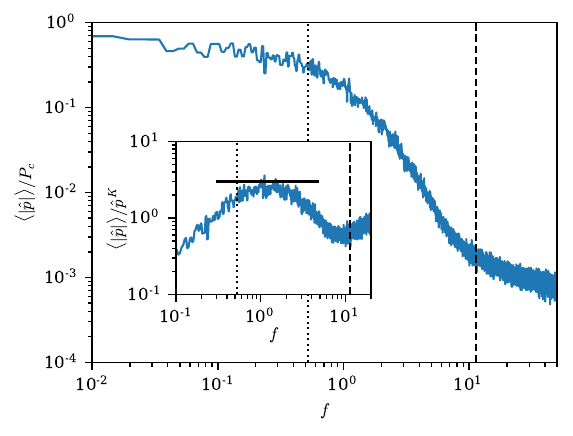}
\caption{a)~Amplitude of the local pressure Fourier transform. The vertical dotted line corresponds to the frequency $f_c(L)$ of eddies of integral length scale in size, while the dashed line corresponds to the frequency $f_c(\eta)$ of eddies of Kolmogorov length scale in size, . Inset plot: Compensated Fourier transform $\avg{|\hat p |} p^{-K}$ with $K = -4/3$. The solid line corresponds to  $\avg{|\hat p |} = 3 p^{K}$.}
\label{fig:spectrum-p-loc}
\end{figure}

To investigate the frequency statistics of the local pressure, we compute the temporal Fourier transform of each pressure signal $\hat p$,
\begin{equation}
    \hat p(\omega) = \int_{-\infty}^\infty p(t) e^{-i \omega t} \dd t.
\end{equation}
The average amplitude of its Fourier transform $\langle |\hat p| \rangle$ is plotted on figure~\ref{fig:spectrum-p-loc}.
The corresponding inertial range in frequency space is delimited by the inverse of the eddy turnover time at the integral scale $f_c(L) = 1/t_c(L)$ (black dotted line) and the inverse of the eddy turnover time at the Kolmogorov scale, $f_c(\eta)$ (dashed line). For low frequencies, $f<f_c(L)$, $\langle |\hat p| \rangle$ slowly decreases with $f$.
\citet{abry1994} have shown that this evolution at low frequencies originates from the contribution of vorticity filaments, since their typical lifetime is the integral timescale \citep{douady1991,pumir1994}. Removing their contributions flattens the low frequency spectrum \citep{abry1994}.

In the inertial range of the turbulent cascade, $f_c(L)<f<f_c(\eta)$, $\langle|\hat p|\rangle$ decays down to the noise level near $f_c(\eta)$. In the spatial Fourier space, and \textit{a fortiori} in the temporal Fourier space, there is no consensus for the scaling of the pressure power spectrum within the inertial range \citep{pullin1994}.
A Kolmogorov-like scaling predicts $|\hat p(k)|^2 \sim \epsilon^{4/3} k^{-7/3}$ (reported by \citet{ishihara2003} for instance) but other authors have also reported a $k^{-5/3}$ scaling \citep{vedula1999,gotoh1999}. To transform the spatial power spectrum into a temporal power spectrum, a classical way is to consider that the small scale structures are advected by the large scales.
This is the sweeping hypothesis \citep{kraichnan1964,tennekes1975}, which has been successfully used to reproduce pressure temporal autocorrelation \citep{yao2008}.
Combining this argument with the Kolmogorov prediction, we find that $\langle \hat p \rangle$ should scale as $\hat p^K \sim \epsilon^{2/3} u_{rms}^{5/6} \omega^{-4/3} $, with a proportionality constant of order 1. We find a reasonable agreement, as shown by the compensated spectrum $\langle \hat p \rangle/\hat p^K$ in the inset of figure~\ref{fig:spectrum-p-loc}. As evidenced by \citet{pumir1994,pullin1994,vedula1999} the Kolmogorov scaling might only hold in a narrow range of frequencies, corresponding to scales just below the integral scale, due to the limited inertial range. The proportionality constant is around 3 in our case (solid black line) lower than the value of 7 proposed by \citet{pumir1994}. The third regime $f>f_c(\eta)$, corresponds to the end of the inertial range and is close to the limit of resolution of our DNS, as $f_c({\Delta x}) = 3 f_c(\eta)$, where $\Delta x$ is the minimum grid size. 

\subsection{Pressure field on a sphere}
\label{subsec:pressuremodes}

\begin{figure}
\centering
\includegraphics{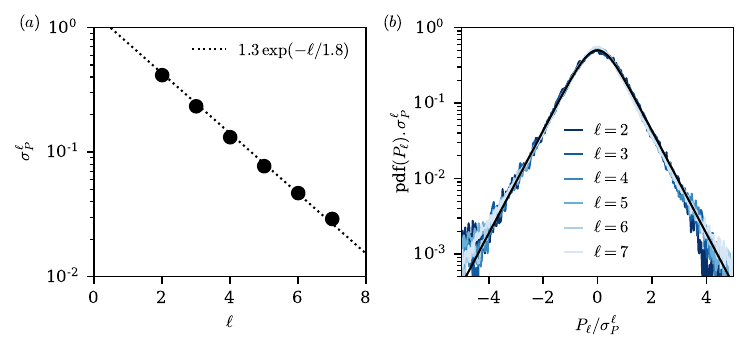}
\caption{a)~Pressure standard deviation, $\sigma_P^\ell$ as a function of $\ell$, showing an exponential decay with $\ell$ (black dotted line) b)~Distributions of $P_\ell$, normalized by $\sigma_P^\ell$, as a function of $\ell$. All the pressure modes share the forcing distribution given in equation~\eqref{eq:distribthforce}.}
\label{fig:std-pdf-pressure}
\end{figure}

To compare the pressure forcing with the effective forcing $\force_\ell$, we interpolate the pressure field $p_S(\theta,\phi)$ in the single phase DNS on a sphere of radius $\R$, and compute its spherical harmonics decomposition
\begin{equation}
    p_S(\theta, \phi) = P_c \big[\sum_{\ell = 0}^\infty \sum_{m = -\ell}^\ell P_{\ell, m}(t)\,Y_\ell^m(\theta, \phi)\big].
\end{equation}
Similarly to the modes of deformation $x_{\ell, m}$, the statistics of $P_{\ell, m}$ are independent of $m$. Ensemble averages are then computed over the three simulations and the $m$ values. For pressure, the modes $\ell =0$ and $\ell = 1$ are non zero, however we focus in the following on modes $\ell\geq 2$ which are relevant for bubble deformations. Figure~\ref{fig:std-pdf-pressure}a shows that the standard deviation of each mode $\ell$, $\sigma_P^\ell$, varies exponentially with $\ell$. A higher $\ell$ is associated to fluctuations at a smaller scale, which are known to be less energetic. However we have no explanation for the exponential scaling. We also observed a decay of $\sigma_\force^\ell$ with $\ell$ (figure~\ref{fig:stdforce}).
The symmetry between positive and negative values is restored, as shown on figure~\ref{fig:std-pdf-pressure}b. Distributions now show exponential tails for both negative and positive pressure values. The shape of the distribution is found independent of $\ell$, corresponding to the same hyperbolic secant distribution (eq.~\eqref{eq:distribthforce}) than the effective forcing distribution we previously identified.

\begin{figure}
\centering
\includegraphics{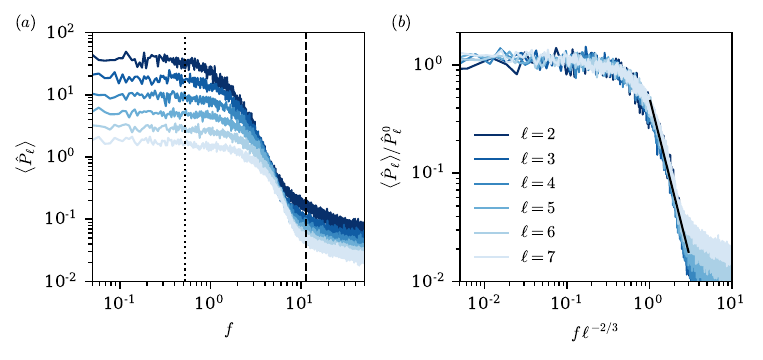}
\caption{a)~Amplitude of the pressure Fourier transform  $\langle |\hat P_\ell| \rangle$ for each mode $\ell$ as a function of the frequency in unit of the eddy turn over time at the bubble scale. The black dashed line is represents the eddy turnover time at scale $\eta$. The black dotted line is the eddy turnover time at the integral length scale. b)~Normalised pressure Fourier transform as a function of frequency in unit of the eddy turnover time at scale $d/\ell$. The black line follows $f^{-3}$.}
\label{fig:spectrum-pressure}
\end{figure}

Eventually, we compute the temporal Fourier transform $\hat P_\ell$ of the spherical pressure modes $P_\ell$. Figure~\ref{fig:spectrum-pressure}a shows the ensemble average of the norm, $\langle |\hat P_\ell| \rangle$ as a function of the frequency. 
For each $\ell$, we recover the three regimes we observed for the point pressure spectrum and $\force_\ell$. 
The transition between the two first regimes depends on the mode $\ell$. Considering that the pressure spectrum share the same characteristic frequency than the effective forcing spectrum, we expect the transition to occur at $f = \ell^{2/3}$, the frequency associated with eddies of size $d/\ell$, in units of $t_c(d)$. 
We show in figure~\ref{fig:spectrum-pressure}b the spectra $\langle |\hat P_\ell| \rangle$ normalized by their low frequency limit, $\hat P_\ell^0$, as a function of the frequency normalized by $\ell^{2/3}$, the eddy turnover time at scale $d/\ell$. 
All curves collapse on a single master curve, showing that pressure and effective forcing share the same time scales. Below the critical frequency ($f<f_\ell$), the spectrum amplitude converges to a constant value, significantly above the integral frequency $f_L$. 
Similarly to \citet{abry1994}, the pressure spectrum at low frequency is now constant. We can assume that the averaging over the sphere has filtered the contribution from localized structures, and in particular the vorticity filaments. A flat spectrum in the range $f_c(L)<f<\ell^{2/3}$ also indicates that the contribution of eddies larger than $d/\ell$, which are roughly homogeneous at the mode scale, has also been filtered out: a bubble is mainly deformed by eddies at its scale. For $\ell^{2/3}<f<f_\eta$, $\langle |\hat P_\ell| \rangle$ follows $f^{-3}$. This decay is steeper than the $\ell$-dependency of $\langle |\hat\force_\ell| \rangle$ which follows $f^{-2}$. This might be attributed to the discrepancy between Eulerian and Lagrangian statistics. From sweeping effect \citep{kraichnan1964}, the temporal decorrelation of Eulerian quantities are expected to occur faster than their Lagrangian counterpart.

To summarize, we have shown that the effective forcing $\force_\ell$ deforming a bubble inherits the pdf of pressure modes integrated over a sphere of same radius. As a consequence of the filtering effect induced by the integration over a sphere, the characteristic frequency associated to each mode $\ell$ is the eddy turnover time at scale $d/\ell$, the frequencies smaller than $\ell^{2/3}$ are well described by a white noise, and the forcing amplitude decreases with $\ell$.

\subsection{Dissipation profiles}

Our analysis of bubble deformation show that (i)~The effective forcing originates from pressure fluctuations near the bubble, and it is not affected by bubble deformability. (ii)~The damping of bubble oscillations is significantly enhanced compare to the quiescent case. We interpreted this additional damping by the presence of a boundary layer of size $L_t = \R/5$ independent of We.
In this section we investigate the velocity gradient profile near the bubble, on a distance comparable to bubble typical deformation. To do so, we need to compute a local distance $r$ to the interface, which is not provided by the Basilisk VOF algorithm.

The method principle is the following. For every bulk point, we find the closest grid point on the interface. We then interpolate the bubble surface around this point, using a quadratic interpolation on the $20$ closest neighbouring interfacial points. To find the neighbours efficiently, the interfacial grid points are stored in a k-d tree structure. The distance $r$ to the interface is then found by minimizing the distance from the bulk point to the quadratic manifold. We follow this procedure for both outside ($r>0$) and inside ($r<0$) bulk points.

\begin{figure}
\centering
\includegraphics{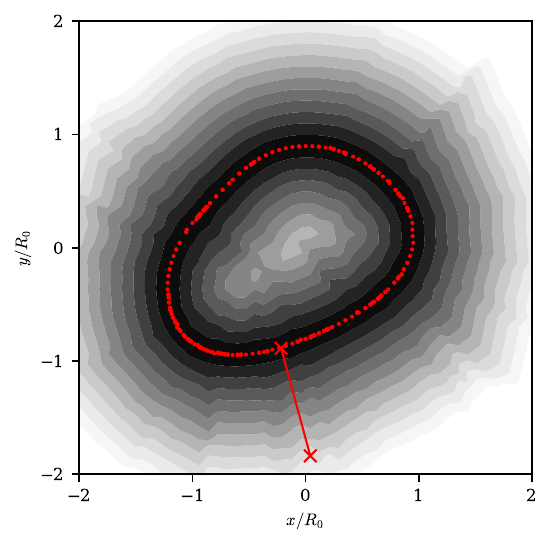}
\caption{Example of the distance computation on a slice of a bubble at $\We = 2$. The red points are on the interface. Isocontours are separated by $0.0625\R$. We also show the association between one bulk points and the corresponding interfacial point.}
\label{fig:schemedistance}
\end{figure}

We diagnose the additional dissipative term of the linear model by investigating the local dissipation rate profile around the bubble.
The energy dissipation rate per unit of mass in a elementary volume is related to the local velocity gradients by
\begin{equation}
    \avg{\epsilon}(x) = 2 \nu \avg{(\partial_i u_j + \partial_j u_i)^2},
\end{equation}
where we use Einstein notations.
For each run, we output snapshots of the full flow field at times separated by at least one eddy turnover time at the bubble scale, to ensure statistical independence.
We then compute profiles of the local dissipation near the interface by averaging on shells of constant distance from the bubble interface, as illustrated on figure~\ref{fig:schemedistance}. Eventually, for each Weber number, we ensemble average the flow snapshots (see table~\ref{tab:Nbrdumps}) to extract a mean profile.

\begin{figure}
\centering
\includegraphics{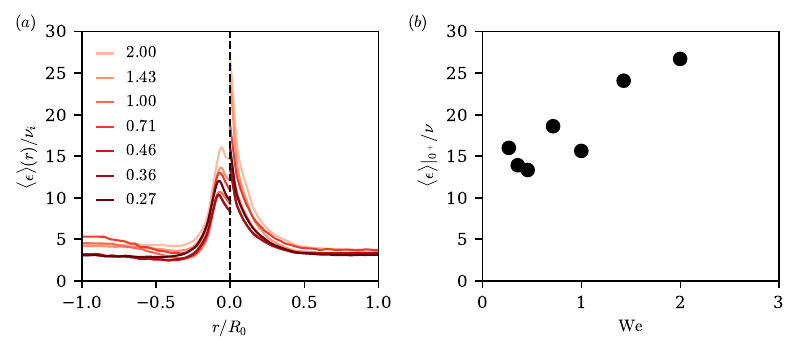}
\caption{a)~Local velocity gradient inside and outside the bubble as a function of the distance to the interface, for all We. b)~Limit of the dissipation rate at the bubble interface in the liquid phase as a function of the Weber number.}
\label{fig:profiles}
\end{figure}

Figure~\ref{fig:profiles}a shows the average local dissipation, divided by the kinematic viscosity, $\avg{\epsilon}(r)/\nu$, as a function of the distance $r$ to the bubble interface.
Far from the bubble interface, for $r>\R/2$ and $r<-\R/2$, the local dissipation converges to a constant.
In the gas, velocity gradients are maximum at $r=-\R/15$. The existence of a maximum inside the bubble near the interface originates from the nearly no slip boundary condition imposed by the denser fluid on the gas inside the bubble.
Similar boundary layer has indeed been observed near solid particle surface (no slip boundary condition)~\citep{shen2022,chiarini2024}.
For bubbles, we therefore expect that decreasing the gas density increases the amplitude of the peak.
The velocity gradients inside and outside the bubble share the same order of magnitude: the dissipation hence mainly takes place outside the bubble, in the liquid, where the dynamical viscosity is much larger. To understand the origin of the additional dissipation we then focus on the outside boundary layer.

\begin{table}
    \centering
    \begin{tabular}{|c | c c c c c c c |} 
    \hline
    We & 2 & 1.43 & 1 & 0.71 & 0.46 & 0.36 & 0.27 \\ [0.5ex] 
     \hline
     N & 48 & 68 & 68 & 27 & 46 & 24 & 52\\ [0.5ex]
     \hline
     \end{tabular}   
     \caption{Number of snapshots per Weber number used to compute the flow profiles.}
    \label{tab:Nbrdumps}
\end{table}

For $r>0$, we observe a thick boundary layer of typical size $\R/5$ (see figure~\ref{fig:profiles}b), compatible with our estimation of $L_t$ (see equation~\eqref{eq:Ldiss}).
Figure~\ref{fig:profiles}b shows the dissipation rate value at the  interface, in the liquid $\avg{\epsilon}|_{0^+}$ as a function of the Weber number.
At vanishing Weber number, we find a non zero dissipation originating from a geometrical boundary layer.
The interfacial value varies between 3 times ($\We = 0.27$) and four times ($\We = 2$) larger than in the bulk.
In addition, we find an increase compatible with a linear dependency of the interfacial dissipation with We.
If we interpret this additional dissipation as an energy transfer rate from the surface deformation to the flow, it would scale as $\Lambda_\ell (\dot x_\ell)^2$.
We have $(\dot x_\ell)^2 \sim (\omega_\ell \sigma_x^{\ell})^2 \propto \We$.
This interpretation is therefore compatible with a damping coefficient $\Lambda$ independent of $\We$.

In the absence of flow, the thickness of the boundary layer of the oscillating bubble can be estimated by $\sqrt{2\nu/\omega_2}$.
For a Weber number ranging from 2.9 to 0.27, this estimation gives a boundary layer of size ranging from $0.07\R$ to $0.04\R$, which is much thinner than the boundary layer thickness we measured. We conclude that the boundary layer originates from a geometrical turbulent boundary layer, and not from bubble oscillations. The existence of a thick boundary layer was completely disregarded in the computation of \citet{lamb1932} for a potential flow far from the interface. The thick boundary layer we observed for the dissipation profile is then consistent with a viscous damping one order of magnitude larger than in the quiescent case.

\section{Conclusion}
In summary, we have shown that the deformations of bubbles in turbulence can be described in terms of a stochastic linear oscillator on the Rayleigh modes of oscillations up to a Weber of order unity. 
Conversely to previous works, we have directly measured using DNS of bubbles in turbulence the coefficients of this reduced model, namely, the damping rate and the natural frequency, together with the statistical properties of the effective forcing.
We have shown that the natural frequency associated to each mode of deformation is not modified compared to the quiescent case. For the effective dissipation, we found that the damping is one order of magnitude larger than the prediction from Lamb. Looking at the dissipation profiles near the interface, we confirmed that the additional dissipation originates from a thick geometrical boundary layer of size $L_t \approx \R/5$ in our case.
In physical units, we expect the damping coefficients $\Lambda_\ell$ to scale as $\nu/d^2 \Re(d)$.
Eventually, we found that the effective forcing, which results from the integration of pressure modes on the bubble surface, does not depend on the Weber number.
This observation confirms that bubble deformations are one-way coupled to the flow: the back-reaction of bubble deformations on the surrounding turbulent flow can be neglected. This effective forcing is characterized by a probability distribution with exponential tails and a typical correlation time which scales with the eddy turnover time at the mode's scale $t_c(d/\ell)$. We also looked at the statistics of pressure fluctuations on a sphere in the absence of bubbles, and we showed that the effective forcing shares the same pdf as the pressure modes' pdf as well as the same characteristic timescale.
Due to the enhanced dissipation compared to the quiescent case, we showed that the resonant oscillation mechanism is not statistically relevant to explain break-ups. Indeed, at Weber of order unity, the bubble cannot accumulate deformation energy on several periods of oscillations as the quality factor $Q = \omega_2/\lambda$ of the main bubble oscillations is too small. Instead, the critical Weber number is set by the interaction between the statistics of large pressure fluctuations and bubble deformations.

\begin{acknowledgements}
We thank Adrian Van Kan, Luc Deike and François P\'etr\'elis for scientific discussions. We also thank Christophe Josserand and Laurent Duchemin for fruitful discussions. This work was performed using HPC resources from GENCI-IDRIS (Grant 2023-AD012B14107). This work was also granted access to the HPC resources of MesoPSL financed by the Region Ile de France and the project Equip@Meso (reference
ANR-10-EQPX-29-01) of the programme Investissements d'Avenir supervised by the Agence Nationale pour la Recherche. This work was funded by a PSL Starting Grant 2022, as well as the ANR Lascaturb (reference ANR-23-CE30-0043-03). Declaration of interests. The authors declare no conflict of interest.
\end{acknowledgements}

\bibliographystyle{jfm}
\bibliography{biblio}

\begin{thebibliography}{78}
\expandafter\ifx\csname natexlab\endcsname\relax\def\natexlab#1{#1}\fi
\def\au#1{#1} \def\ed#1{#1} \def\yr#1{#1}\def\at#1{#1}\def\jt#1{\textit{#1}}
  \def\bt#1{#1}\def\bvol#1{\textbf{#1}} \def\vol#1{#1} \def\pg#1{#1}
  \def\publ#1{#1}\def\arxiv#1{#1}\def\org#1{#1}\def\st#1{\textit{#1}}

\bibitem[Abry {\em et~al.\/}(1994)Abry, Fauve, Flandrin \& Laroche]{abry1994}
{\sc \au{Abry, Patrice}, \au{Fauve, S}, \au{Flandrin, P} \& \au{Laroche, C}}
  \yr{1994}  \at{Analysis of pressure fluctuations in swirling turbulent
  flows}.  \jt{Journal de Physique II}  \bvol{4}~(5),  \pg{725--733}.

\bibitem[Abu-Al-Saud {\em et~al.\/}(2018)Abu-Al-Saud, Popinet \&
  Tchelepi]{basiliksurfacetension}
{\sc \au{Abu-Al-Saud, Moataz~O.}, \au{Popinet, St{\'e}phane} \& \au{Tchelepi,
  Hamdi~A.}} \yr{2018}  \at{{A conservative and well-balanced surface tension
  model}}.  \jt{{Journal of Computational Physics}} .

\bibitem[Bappy {\em et~al.\/}(2019)Bappy, Carrica \& Buscaglia]{bappy2019}
{\sc \au{Bappy, Mehedi}, \au{Carrica, Pablo~M} \& \au{Buscaglia, Gustavo~C}}
  \yr{2019}  \at{Lagrangian statistics of pressure fluctuation events in
  homogeneous isotropic turbulence}.  \jt{Physics of Fluids}  \bvol{31}~(8).

\bibitem[Bappy {\em et~al.\/}(2020{\natexlab{{\em a\/}}})Bappy, Carrica,
  Vela-Mart{\'\i}n, Freire \& Buscaglia]{bappy2020pressure}
{\sc \au{Bappy, Mehedi~H}, \au{Carrica, Pablo~M}, \au{Vela-Mart{\'\i}n,
  Alberto}, \au{Freire, Livia~S} \& \au{Buscaglia, Gustavo~C}}
  \yr{2020{\natexlab{{\em a\/}}}}  \at{Pressure statistics of gas nuclei in
  homogeneous isotropic turbulence with an application to cavitation
  inception}.  \jt{Physics of Fluids}  \bvol{32}~(9).

\bibitem[Bappy {\em et~al.\/}(2020{\natexlab{{\em b\/}}})Bappy, Vela-Martin,
  Buscaglia, Carrica \& Freire]{bappy2020effect}
{\sc \au{Bappy, Mehedi~Hasan}, \au{Vela-Martin, Alberto}, \au{Buscaglia,
  Gustavo~Carlos}, \au{Carrica, Pablo~M} \& \au{Freire, L{\'\i}via~Souza}}
  \yr{2020{\natexlab{{\em b\/}}}} Effect of bubble size on lagrangian pressure
  statistics in homogeneous isotropic turbulence.  \bt{In {\em Journal of
  Physics: Conference Series\/}}, ,  \vol{vol. 1522},  \pg{p. 012002}. IOP
  Publishing.

\bibitem[Berry {\em et~al.\/}(2005)Berry, Hyers, Racz \& Abedian]{berry2005}
{\sc \au{Berry, SR}, \au{Hyers, RW}, \au{Racz, LM} \& \au{Abedian, B}}
  \yr{2005}  \at{Surface oscillations of an electromagnetically levitated
  droplet}.  \jt{International journal of thermophysics}  \bvol{26},
  \pg{1565--1581}.

\bibitem[Bojarevics \& Pericleous(2003)]{bojarevics2003}
{\sc \au{Bojarevics, Valdis} \& \au{Pericleous, Koulis}} \yr{2003}
  \at{Modelling electromagnetically levitated liquid droplet oscillations}.
  \jt{ISIJ international}  \bvol{43}~(6),  \pg{890--898}.

\bibitem[Boussinesq(1877)]{boussinesq1877}
{\sc \au{Boussinesq, Joseph}} \yr{1877} {\em Essai sur la th{\'e}orie des eaux
  courantes\/}.  \publ{Impr. nationale}.

\bibitem[Brouzet {\em et~al.\/}(2021)Brouzet, Guin{\'e}, Dalbe, Favier,
  Vandenberghe, Villermaux \& Verhille]{brouzet2021}
{\sc \au{Brouzet, Christophe}, \au{Guin{\'e}, Rapha{\"e}l}, \au{Dalbe,
  Marie-Julie}, \au{Favier, Benjamin}, \au{Vandenberghe, Nicolas},
  \au{Villermaux, Emmanuel} \& \au{Verhille, Gautier}} \yr{2021}
  \at{Laboratory model for plastic fragmentation in the turbulent ocean}.
  \jt{Physical Review Fluids}  \bvol{6}~(2),  \pg{024601}.

\bibitem[Cadot {\em et~al.\/}(1995)Cadot, Douady \& Couder]{cadot1995}
{\sc \au{Cadot, O}, \au{Douady, S} \& \au{Couder, Y}} \yr{1995}
  \at{Characterization of the low-pressure filaments in a three-dimensional
  turbulent shear flow}.  \jt{Physics of Fluids}  \bvol{7}~(3),  \pg{630--646}.

\bibitem[Calzavarini {\em et~al.\/}(2009)Calzavarini, Volk, Bourgoin,
  L{\'e}v{\^e}que, Pinton \& Toschi]{calzavarini2009}
{\sc \au{Calzavarini, Enrico}, \au{Volk, Romain}, \au{Bourgoin, Micka{\"e}l},
  \au{L{\'e}v{\^e}que, Emmanuel}, \au{Pinton, J-F} \& \au{Toschi, Federico}}
  \yr{2009}  \at{Acceleration statistics of finite-sized particles in turbulent
  flow: the role of fax{\'e}n forces}.  \jt{Journal of Fluid Mechanics}
  \bvol{630},  \pg{179--189}.

\bibitem[Cao {\em et~al.\/}(1999)Cao, Chen \& Doolen]{cao1999}
{\sc \au{Cao, Nianzheng}, \au{Chen, Shiyi} \& \au{Doolen, Gary~D}} \yr{1999}
  \at{Statistics and structures of pressure in isotropic turbulence}.
  \jt{Physics of Fluids}  \bvol{11}~(8),  \pg{2235--2250}.

\bibitem[Chandrasekhar(1959)]{chandrasekhar1959}
{\sc \au{Chandrasekhar, Subrahmanyan}} \yr{1959}  \at{The oscillations of a
  viscous liquid globe}.  \jt{Proceedings of the London Mathematical Society}
  \bvol{3}~(1),  \pg{141--149}.

\bibitem[Chandrasekhar(2013)]{chandrasekhar2013}
{\sc \au{Chandrasekhar, Subrahmanyan}} \yr{2013} {\em Hydrodynamic and
  hydromagnetic stability\/}.  \publ{Courier Corporation}.

\bibitem[Chiarini \& Rosti(2024)]{chiarini2024}
{\sc \au{Chiarini, Alessandro} \& \au{Rosti, Marco~Edoardo}} \yr{2024}
  \at{Finite-size inertial spherical particles in turbulence}.  \jt{arXiv
  preprint arXiv:2404.16475} .

\bibitem[De~Langre(2008)]{langre2008}
{\sc \au{De~Langre, Emmanuel}} \yr{2008}  \at{Effects of wind on plants}.
  \jt{Annu. Rev. Fluid Mech.}  \bvol{40},  \pg{141--168}.

\bibitem[Deike(2022)]{deike2022}
{\sc \au{Deike, Luc}} \yr{2022}  \at{Mass transfer at the ocean--atmosphere
  interface: the role of wave breaking, droplets, and bubbles}.  \jt{Annual
  Review of Fluid Mechanics}  \bvol{54},  \pg{191--224}.

\bibitem[Douady {\em et~al.\/}(1991)Douady, Couder \& Brachet]{douady1991}
{\sc \au{Douady, S}, \au{Couder, Y} \& \au{Brachet, ME}} \yr{1991}  \at{Direct
  observation of the intermittency of intense vorticity filaments in
  turbulence}.  \jt{Physical review letters}  \bvol{67}~(8),  \pg{983}.

\bibitem[Fan {\em et~al.\/}(2024)Fan, Wang, Jiang, Sun \& Calzavarini]{fan2024}
{\sc \au{Fan, Yaning}, \au{Wang, Cheng}, \au{Jiang, Linfeng}, \au{Sun, Chao} \&
  \au{Calzavarini, Enrico}} \yr{2024}  \at{Accelerations of large inertial
  particles in turbulence}.  \jt{Europhysics Letters} .

\bibitem[Fauve {\em et~al.\/}(1993)Fauve, Laroche \& Castaing]{fauve1993}
{\sc \au{Fauve, Stephan}, \au{Laroche, C} \& \au{Castaing, B}} \yr{1993}
  \at{Pressure fluctuations in swirling turbulent flows}.  \jt{Journal de
  physique II}  \bvol{3}~(3),  \pg{271--278}.

\bibitem[Galinat {\em et~al.\/}(2007)Galinat, Risso, Masbernat \&
  Guiraud]{galinat2007}
{\sc \au{Galinat, Sophie}, \au{Risso, Fr{\'e}d{\'e}ric}, \au{Masbernat,
  Olivier} \& \au{Guiraud, Pascal}} \yr{2007}  \at{Dynamics of drop breakup in
  inhomogeneous turbulence at various volume fractions}.  \jt{Journal of Fluid
  Mechanics}  \bvol{578},  \pg{85--94}.

\bibitem[Gitterman(2005)]{gitterman2005}
{\sc \au{Gitterman, Moshe}} \yr{2005} {\em Noisy Oscillator, The: The First
  Hundred Years, From Einstein Until Now\/}.  \publ{World Scientific}.

\bibitem[Gotoh \& Rogallo(1999)]{gotoh1999}
{\sc \au{Gotoh, Toshiyuki} \& \au{Rogallo, Robert~S}} \yr{1999}
  \at{Intermittency and scaling of pressure at small scales in forced isotropic
  turbulence}.  \jt{Journal of Fluid Mechanics}  \bvol{396},  \pg{257--285}.

\bibitem[H{\aa}kansson(2019)]{haakansson2019}
{\sc \au{H{\aa}kansson, Andreas}} \yr{2019}  \at{Emulsion formation by
  homogenization: Current understanding and future perspectives}.  \jt{Annual
  review of food science and technology}  \bvol{10},  \pg{239--258}.

\bibitem[H{\aa}kansson(2021)]{haakansson2021}
{\sc \au{H{\aa}kansson, Andreas}} \yr{2021}  \at{The role of stochastic
  time-variations in turbulent stresses when predicting drop breakup—a review
  of modelling approaches}.  \jt{Processes}  \bvol{9}~(11),  \pg{1904}.

\bibitem[Hinze(1955)]{hinze1955}
{\sc \au{Hinze, Julius~O}} \yr{1955}  \at{Fundamentals of the hydrodynamic
  mechanism of splitting in dispersion processes}.  \jt{AIChE journal}
  \bvol{1}~(3),  \pg{289--295}.

\bibitem[Homann \& Bec(2010)]{homann2010}
{\sc \au{Homann, Holger} \& \au{Bec, Jeremie}} \yr{2010}  \at{Finite-size
  effects in the dynamics of neutrally buoyant particles in turbulent flow}.
  \jt{Journal of Fluid Mechanics}  \bvol{651},  \pg{81--91}.

\bibitem[Ishihara {\em et~al.\/}(2003)Ishihara, Kaneda, Yokokawa, Itakura \&
  Uno]{ishihara2003}
{\sc \au{Ishihara, Takashi}, \au{Kaneda, Yukio}, \au{Yokokawa, Mitsuo},
  \au{Itakura, Ken'ichi} \& \au{Uno, Atsuya}} \yr{2003}  \at{Spectra of energy
  dissipation, enstrophy and pressure by high-resolution direct numerical
  simulations of turbulence in a periodic box}.  \jt{journal of the physical
  society of japan}  \bvol{72}~(5),  \pg{983--986}.

\bibitem[Kang \& Leal(1988)]{kang1988}
{\sc \au{Kang, IS} \& \au{Leal, LG}} \yr{1988}  \at{Small-amplitude
  perturbations of shape for a nearly spherical bubble in an inviscid straining
  flow (steady shapes and oscillatory motion)}.  \jt{Journal of Fluid
  Mechanics}  \bvol{187},  \pg{231--266}.

\bibitem[Kolmogorov(1949)]{kolmogorov1949}
{\sc \au{Kolmogorov, Andrey}} \yr{1949} On the breakage of drops in a turbulent
  flow.  \bt{In {\em Dokl. Akad. Navk. SSSR\/}}, ,  \vol{vol.~66},  \pg{pp.
  825--828}.

\bibitem[Kraichnan(1964)]{kraichnan1964}
{\sc \au{Kraichnan, Robert~H}} \yr{1964}  \at{Kolmogorov’s hypotheses and
  eulerian turbulence theory}.  \jt{Phys. Fluids}  \bvol{7}~(11),
  \pg{1723--1734}.

\bibitem[Lalanne {\em et~al.\/}(2019)Lalanne, Masbernat \& Risso]{lalanne2019}
{\sc \au{Lalanne, Benjamin}, \au{Masbernat, Olivier} \& \au{Risso,
  Fr{\'e}d{\'e}ric}} \yr{2019}  \at{A model for drop and bubble breakup
  frequency based on turbulence spectra}.  \jt{AIChE Journal}  \bvol{65}~(1),
  \pg{347--359}.

\bibitem[Lamb(1932)]{lamb1932}
{\sc \au{Lamb, Horace}} \yr{1932}  \at{Hydrodynamics}.  \jt{Aufl., Cambridge:
  Univ. Press1879--1932}  \bvol{427}.

\bibitem[Lee {\em et~al.\/}(1987)Lee, Erickson \& Glasgow]{lee1987}
{\sc \au{Lee, Chung-Hur}, \au{Erickson, LE} \& \au{Glasgow, LA}} \yr{1987}
  \at{Bubble breakup and coalescence in turbulent gas-liquid dispersions}.
  \jt{Chemical Engineering Communications}  \bvol{59}~(1-6),  \pg{65--84}.

\bibitem[Luo \& Svendsen(1996)]{luo1996}
{\sc \au{Luo, Hean} \& \au{Svendsen, Hallvard~F}} \yr{1996}  \at{Theoretical
  model for drop and bubble breakup in turbulent dispersions}.  \jt{AIChE
  journal}  \bvol{42}~(5),  \pg{1225--1233}.

\bibitem[Maniero {\em et~al.\/}(2012)Maniero, Masbernat, Climent \&
  Risso]{maniero2012}
{\sc \au{Maniero, Riccardo}, \au{Masbernat, Olivier}, \au{Climent, Eric} \&
  \au{Risso, Fr{\'e}d{\'e}ric}} \yr{2012}  \at{Modeling and simulation of
  inertial drop break-up in a turbulent pipe flow downstream of a restriction}.
   \jt{International journal of multiphase flow}  \bvol{42},  \pg{1--8}.

\bibitem[Mart{\'i}nez-Baz{\'a}n {\em et~al.\/}(1999)Mart{\'i}nez-Baz{\'a}n,
  Montanes \& Lasheras]{martinez1999a}
{\sc \au{Mart{\'i}nez-Baz{\'a}n, Carlos}, \au{Montanes, JL} \& \au{Lasheras,
  Juan~C}} \yr{1999}  \at{On the breakup of an air bubble injected into a fully
  developed turbulent flow. part 1. breakup frequency}.  \jt{Journal of Fluid
  Mechanics}  \bvol{401},  \pg{157--182}.

\bibitem[Masuk {\em et~al.\/}(2021{\natexlab{{\em a\/}}})Masuk, Qi, Salibindla
  \& Ni]{masuk2021model}
{\sc \au{Masuk, Ashik Ullah~Mohammad}, \au{Qi, Yinghe}, \au{Salibindla,
  Ashwanth~KR} \& \au{Ni, Rui}} \yr{2021{\natexlab{{\em a\/}}}}  \at{Towards a
  phenomenological model on the deformation and orientation dynamics of
  finite-sized bubbles in both quiescent and turbulent media}.  \jt{Journal of
  Fluid Mechanics}  \bvol{920},  \pg{A4}.

\bibitem[Masuk {\em et~al.\/}(2021{\natexlab{{\em b\/}}})Masuk, Salibindla \&
  Ni]{masuk2021simultaneous}
{\sc \au{Masuk, Ashik Ullah~Mohammad}, \au{Salibindla, Ashwanth~KR} \& \au{Ni,
  Rui}} \yr{2021{\natexlab{{\em b\/}}}}  \at{Simultaneous measurements of
  deforming hinze-scale bubbles with surrounding turbulence}.  \jt{Journal of
  Fluid Mechanics}  \bvol{910}.

\bibitem[Maxey \& Riley(1983)]{maxey1983}
{\sc \au{Maxey, Martin~R} \& \au{Riley, James~J}} \yr{1983}  \at{Equation of
  motion for a small rigid sphere in a nonuniform flow}.  \jt{The Physics of
  Fluids}  \bvol{26}~(4),  \pg{883--889}.

\bibitem[Miller \& Scriven(1968)]{miller1968}
{\sc \au{Miller, CA} \& \au{Scriven, LE}} \yr{1968}  \at{The oscillations of a
  fluid droplet immersed in another fluid}.  \jt{Journal of fluid mechanics}
  \bvol{32}~(3),  \pg{417--435}.

\bibitem[Mordant {\em et~al.\/}(2004)Mordant, Crawford \&
  Bodenschatz]{mordant2004}
{\sc \au{Mordant, Nicolas}, \au{Crawford, Alice~M} \& \au{Bodenschatz,
  Eberhard}} \yr{2004}  \at{Three-dimensional structure of the lagrangian
  acceleration in turbulent flows}.  \jt{Physical review letters}
  \bvol{93}~(21),  \pg{214501}.

\bibitem[Perrard {\em et~al.\/}(2021)Perrard, Rivi{\`e}re, Mostert \&
  Deike]{perrard2021}
{\sc \au{Perrard, St{\'e}phane}, \au{Rivi{\`e}re, Ali{\'e}nor}, \au{Mostert,
  Wouter} \& \au{Deike, Luc}} \yr{2021}  \at{Bubble deformation by a turbulent
  flow}.  \jt{Journal of Fluid Mechanics}  \bvol{920},  \pg{A15}.

\bibitem[Pope(2000)]{pope2000}
{\sc \au{Pope, S.~B.}} \yr{2000} {\em Turbulent flows\/}.  \publ{Cambridge
  university press, Cambridge, UK}.

\bibitem[Popinet(2003)]{popinet2003}
{\sc \au{Popinet, S.}} \yr{2003}  \at{Gerris: a tree-based adaptive solver for
  the incompressible euler equations in complex geometries}.  \jt{J. Comput.
  Phys.}  \bvol{190}~(2),  \pg{572--600}.

\bibitem[Popinet(2009)]{popinet2009}
{\sc \au{Popinet, S.}} \yr{2009}  \at{An accurate adaptive solver for
  surface-tension-driven interfacial flows}.  \jt{Journal of Computational
  Physics}  \bvol{228},  \pg{5838--5866}.

\bibitem[Prakash {\em et~al.\/}(2012)Prakash, Tagawa, Calzavarini, Mercado,
  Toschi, Lohse \& Sun]{prakash2012}
{\sc \au{Prakash, Vivek~N}, \au{Tagawa, Yoshiyuki}, \au{Calzavarini, Enrico},
  \au{Mercado, Juli{\'a}n~Mart{\'\i}nez}, \au{Toschi, Federico}, \au{Lohse,
  Detlef} \& \au{Sun, Chao}} \yr{2012}  \at{How gravity and size affect the
  acceleration statistics of bubbles in turbulence}.  \jt{New journal of
  physics}  \bvol{14}~(10),  \pg{105017}.

\bibitem[Prandtl(1949)]{prandtl1949}
{\sc \au{Prandtl, Ludwig}} \yr{1949}  \at{Report on investigation of developed
  turbulence}.  \jt{Zeitschrift fuer Angewandte Matematik und Mechanik}
  \bvol{5}~(NACA-TM-1231).

\bibitem[Prosperetti(1977)]{prosperetti1977}
{\sc \au{Prosperetti, Andrea}} \yr{1977}  \at{Viscous effects on perturbed
  spherical flows}.  \jt{Quarterly of Applied mathematics}  \bvol{34}~(4),
  \pg{339--352}.

\bibitem[Prosperetti(1980)]{prosperetti1980}
{\sc \au{Prosperetti, Andrea}} \yr{1980}  \at{Free oscillations of drops and
  bubbles: the initial-value problem}.  \jt{Journal of Fluid Mechanics}
  \bvol{100}~(2),  \pg{333--347}.

\bibitem[Pullin \& Rogallo(1994)]{pullin1994}
{\sc \au{Pullin, DI} \& \au{Rogallo, RS}} \yr{1994}  \at{Pressure and
  higher-order spectra for homogeneous isotropic turbulence}.  \jt{Stanford
  Univ., Studying Turbulence Using Numerical Simulation Databases. 5:
  Proceedings of the 1994 Summer Program} .

\bibitem[Pumir(1994)]{pumir1994}
{\sc \au{Pumir, Alain}} \yr{1994}  \at{A numerical study of pressure
  fluctuations in three-dimensional, incompressible, homogeneous, isotropic
  turbulence}.  \jt{Physics of Fluids}  \bvol{6}~(6),  \pg{2071--2083}.

\bibitem[Qureshi {\em et~al.\/}(2007)Qureshi, Bourgoin, Baudet, Cartellier \&
  Gagne]{qureshi2007}
{\sc \au{Qureshi, Nauman~M}, \au{Bourgoin, Micka{\"e}l}, \au{Baudet,
  Christophe}, \au{Cartellier, Alain} \& \au{Gagne, Yves}} \yr{2007}
  \at{Turbulent transport of material particles: an experimental study of
  finite size effects}.  \jt{Physical review letters}  \bvol{99}~(18),
  \pg{184502}.

\bibitem[Ravelet {\em et~al.\/}(2011)Ravelet, Colin \& Risso]{ravelet2011}
{\sc \au{Ravelet, Florent}, \au{Colin, Catherine} \& \au{Risso,
  Fr{\'e}d{\'e}ric}} \yr{2011}  \at{On the dynamics and breakup of a bubble
  rising in a turbulent flow}.  \jt{Physics of Fluids}  \bvol{23}~(10).

\bibitem[Rayleigh(1879)]{rayleigh1879}
{\sc \au{Rayleigh, Lord}} \yr{1879}  \at{On the capillary phenomena of jets}.
  \jt{Proc. R. Soc. London}  \bvol{29}~(196-199),  \pg{71--97}.

\bibitem[Reid(1960)]{reid1960}
{\sc \au{Reid, William~Hill}} \yr{1960}  \at{The oscillations of a viscous
  liquid drop}.  \jt{Quarterly of Applied Mathematics}  \bvol{18}~(1),
  \pg{86--89}.

\bibitem[Risso(2018)]{risso2018}
{\sc \au{Risso, Fr{\'e}d{\'e}ric}} \yr{2018}  \at{Agitation, mixing, and
  transfers induced by bubbles}.  \jt{Annual Review of Fluid Mechanics}
  \bvol{50},  \pg{25--48}.

\bibitem[Risso \& Fabre(1998)]{risso1998}
{\sc \au{Risso, Fr{\'e}d{\'e}ric} \& \au{Fabre, Jean}} \yr{1998}
  \at{Oscillations and breakup of a bubble immersed in a turbulent field}.
  \jt{Journal of Fluid Mechanics}  \bvol{372},  \pg{323--355}.

\bibitem[Rivi{\`e}re {\em et~al.\/}(2023)Rivi{\`e}re, Duchemin, Josserand \&
  Perrard]{riviere2023}
{\sc \au{Rivi{\`e}re, Ali{\'e}nor}, \au{Duchemin, Laurent}, \au{Josserand,
  Christophe} \& \au{Perrard, St{\'e}phane}} \yr{2023}  \at{Bubble breakup
  reduced to a one-dimensional nonlinear oscillator}.  \jt{Physical Review
  Fluids}  \bvol{8}~(9),  \pg{094004}.

\bibitem[Rivi{\`e}re {\em et~al.\/}(2021)Rivi{\`e}re, Mostert, Perrard \&
  Deike]{riviere2021}
{\sc \au{Rivi{\`e}re, Ali{\'e}nor}, \au{Mostert, Wouter}, \au{Perrard,
  St{\'e}phane} \& \au{Deike, Luc}} \yr{2021}  \at{Sub-hinze scale bubble
  production in turbulent bubble break-up}.  \jt{Journal of Fluid Mechanics}
  \bvol{917},  \pg{A40}.

\bibitem[Roa {\em et~al.\/}(2023)Roa, Renoult, Dumouchel \& Br{\"a}ndle~de
  Motta]{roa2023}
{\sc \au{Roa, Ignacio}, \au{Renoult, Marie-Charlotte}, \au{Dumouchel,
  Christophe} \& \au{Br{\"a}ndle~de Motta, Jorge~C{\'e}sar}} \yr{2023}
  \at{Droplet oscillations in a turbulent flow}.  \jt{Frontiers in Physics}
  \bvol{11},  \pg{1173521}.

\bibitem[Rosales \& Meneveau(2005)]{rosales2005}
{\sc \au{Rosales, Carlos} \& \au{Meneveau, Charles}} \yr{2005}  \at{Linear
  forcing in numerical simulations of isotropic turbulence: Physical space
  implementations and convergence properties}.  \jt{Physics of fluids}
  \bvol{17}~(9).

\bibitem[Rosti {\em et~al.\/}(2018)Rosti, Banaei, Brandt \& Mazzino]{rosti2018}
{\sc \au{Rosti, Marco~Edoardo}, \au{Banaei, Arash~Alizad}, \au{Brandt, Luca} \&
  \au{Mazzino, Andrea}} \yr{2018}  \at{Flexible fiber reveals the two-point
  statistical properties of turbulence}.  \jt{Physical review letters}
  \bvol{121}~(4),  \pg{044501}.

\bibitem[Ruth {\em et~al.\/}(2019)Ruth, Mostert, Perrard \& Deike]{ruth2019}
{\sc \au{Ruth, Daniel~J}, \au{Mostert, Wouter}, \au{Perrard, St{\'e}phane} \&
  \au{Deike, Luc}} \yr{2019}  \at{Bubble pinch-off in turbulence}.
  \jt{Proceedings of the National Academy of Sciences}  \bvol{116}~(51),
  \pg{25412--25417}.

\bibitem[Salibindla {\em et~al.\/}(2021)Salibindla, Masuk \&
  Ni]{salibindla2021}
{\sc \au{Salibindla, Ashwanth~KR}, \au{Masuk, Ashik Ullah~Mohammad} \& \au{Ni,
  Rui}} \yr{2021}  \at{Experimental investigation of the acceleration
  statistics and added-mass force of deformable bubbles in intense turbulence}.
   \jt{Journal of Fluid Mechanics}  \bvol{912},  \pg{A50}.

\bibitem[Sevik \& Park(1973)]{sevik1973}
{\sc \au{Sevik, M.} \& \au{Park, S.~H.}} \yr{1973}  \at{{The Splitting of Drops
  and Bubbles by Turbulent Fluid Flow}}.  \jt{Journal of Fluids Engineering}
  \bvol{95}~(1),  \pg{53--60}.

\bibitem[Shen {\em et~al.\/}(2022)Shen, Peng, Wu, Chong, Lu \& Wang]{shen2022}
{\sc \au{Shen, Jie}, \au{Peng, Cheng}, \au{Wu, Jianzhao}, \au{Chong,
  Kai~Leong}, \au{Lu, Zhiming} \& \au{Wang, Lian-Ping}} \yr{2022}
  \at{Turbulence modulation by finite-size particles of different diameters and
  particle--fluid density ratios in homogeneous isotropic turbulence}.
  \jt{Journal of Turbulence}  \bvol{23}~(8),  \pg{433--453}.

\bibitem[Tennekes(1975)]{tennekes1975}
{\sc \au{Tennekes, Henk}} \yr{1975}  \at{Eulerian and lagrangian time
  microscales in isotropic turbulence}.  \jt{Journal of Fluid Mechanics}
  \bvol{67}~(3),  \pg{561--567}.

\bibitem[Toschi \& Bodenschatz(2009)]{toschi2009}
{\sc \au{Toschi, Federico} \& \au{Bodenschatz, Eberhard}} \yr{2009}
  \at{Lagrangian properties of particles in turbulence}.  \jt{Annual review of
  fluid mechanics}  \bvol{41},  \pg{375--404}.

\bibitem[Vedula \& Yeung(1999)]{vedula1999}
{\sc \au{Vedula, Prakash} \& \au{Yeung, Pui-Kuen}} \yr{1999}  \at{Similarity
  scaling of acceleration and pressure statistics in numerical simulations of
  isotropic turbulence}.  \jt{Physics of Fluids}  \bvol{11}~(5),
  \pg{1208--1220}.

\bibitem[Vela-Mart{\'\i}n \& Avila(2021)]{vela2021}
{\sc \au{Vela-Mart{\'\i}n, Alberto} \& \au{Avila, Marc}} \yr{2021}
  \at{Deformation of drops by outer eddies in turbulence}.  \jt{Journal of
  Fluid Mechanics}  \bvol{929},  \pg{A38}.

\bibitem[Verhille(2022)]{verhille2022}
{\sc \au{Verhille, Gautier}} \yr{2022}  \at{Deformability of discs in
  turbulence}.  \jt{Journal of Fluid Mechanics}  \bvol{933},  \pg{A3}.

\bibitem[Volk {\em et~al.\/}(2011)Volk, Calzavarini, Leveque \&
  Pinton]{volk2011}
{\sc \au{Volk, Romain}, \au{Calzavarini, Enrico}, \au{Leveque, Emmanuel} \&
  \au{Pinton, J-F}} \yr{2011}  \at{Dynamics of inertial particles in a
  turbulent von k{\'a}rm{\'a}n flow}.  \jt{Journal of Fluid Mechanics}
  \bvol{668},  \pg{223--235}.

\bibitem[Volk {\em et~al.\/}(2008)Volk, Calzavarini, Verhille, Lohse, Mordant,
  Pinton \& Toschi]{volk2008}
{\sc \au{Volk, R}, \au{Calzavarini, Enrico}, \au{Verhille, G}, \au{Lohse,
  Detlef}, \au{Mordant, N}, \au{Pinton, J-F} \& \au{Toschi, Federico}}
  \yr{2008}  \at{Acceleration of heavy and light particles in turbulence:
  comparison between experiments and direct numerical simulations}.
  \jt{Physica D: Nonlinear Phenomena}  \bvol{237}~(14-17),  \pg{2084--2089}.

\bibitem[Voth {\em et~al.\/}(2002)Voth, La~Porta, Crawford, Alexander \&
  Bodenschatz]{voth2002}
{\sc \au{Voth, Greg~A}, \au{La~Porta, Arthur}, \au{Crawford, Alice~M},
  \au{Alexander, Jim} \& \au{Bodenschatz, Eberhard}} \yr{2002}  \at{Measurement
  of particle accelerations in fully developed turbulence}.  \jt{Journal of
  Fluid Mechanics}  \bvol{469},  \pg{121--160}.

\bibitem[Wang {\em et~al.\/}(2003)Wang, Wang \& Jin]{wang2003}
{\sc \au{Wang, Tiefeng}, \au{Wang, Jinfu} \& \au{Jin, Yong}} \yr{2003}  \at{A
  novel theoretical breakup kernel function for bubbles/droplets in a turbulent
  flow}.  \jt{Chemical Engineering Science}  \bvol{58}~(20),  \pg{4629--4637}.

\bibitem[Xiao {\em et~al.\/}(2021)Xiao, Brillo, Lee, Hyers \& Matson]{xiao2021}
{\sc \au{Xiao, Xiao}, \au{Brillo, J{\"u}rgen}, \au{Lee, Jonghyun}, \au{Hyers,
  Robert~W} \& \au{Matson, Douglas~M}} \yr{2021}  \at{Impact of convection on
  the damping of an oscillating droplet during viscosity measurement using the
  iss-eml facility}.  \jt{npj Microgravity}  \bvol{7}~(1),  \pg{36}.

\bibitem[Yao {\em et~al.\/}(2008)Yao, He, Wang \& Zhang]{yao2008}
{\sc \au{Yao, Hua-Dong}, \au{He, Guo-Wei}, \au{Wang, Meng} \& \au{Zhang, Xing}}
  \yr{2008}  \at{Time correlations of pressure in isotropic turbulence}.
  \jt{Physics of Fluids}  \bvol{20}~(2).

\end{thebibliography}
\end{document}